\documentclass[aps,prl,twocolumn,superscriptaddress,amsmath,amsfonts,amssymb,floatfix,notitlepage,longbibliography]{revtex4-2}

\usepackage{CJK}
\usepackage{soul}
\usepackage[caption=false,subrefformat=parens,labelformat=parens]{subfig}
\usepackage{color}
\usepackage[table,dvipsnames]{xcolor}
\usepackage[colorlinks,linktocpage,bookmarks=false]{hyperref}
\usepackage[normalem]{ulem} 
\usepackage{physics}
\usepackage{natbib}

\newcommand\myshade{85}
\definecolor{myrulecolor}{RGB}{150,20,0}
\colorlet{mylinkcolor}{violet}
\colorlet{mycitecolor}{YellowOrange}
\colorlet{myurlcolor}{Aquamarine}
\hypersetup{
	linkcolor  = myurlcolor!\myshade!black,
	citecolor  = mycitecolor!\myshade!black,
	urlcolor   =  myrulecolor!\myshade!black,
}

\usepackage{graphicx}
\graphicspath{{figure_v4/}{figures/}{figs_sm/}{figure_v2/}}
\usepackage{multirow}
\usepackage{booktabs} 

\AtBeginDocument{
	\heavyrulewidth=.08em
	\lightrulewidth=.05em
	\cmidrulewidth=.03em
	\belowrulesep=.65ex
	\belowbottomsep=0pt
	\aboverulesep=.4ex
	\abovetopsep=0pt
	\cmidrulesep=\doublerulesep
	\cmidrulekern=.5em
	\defaultaddspace=.5em
}

\usepackage[percent]{overpic}

\newcommand{\prlsection}[1]{\noindent\textbf{{#1} --- }}

\renewcommand\[{\begin{equation}}
\renewcommand\]{\end{equation}}

\newcommand{\thistitle}{Signatures of field-induced multi-color kagome spin liquids in the dipole-octupole pyrochlore $\mathrm{Ce_2Hf_2O_7}$}

\begin{document}
	\begin{CJK*}{UTF8}{gbsn} 
		\title{\thistitle}

\author{Edwin Kermarrec}
\thanks{These authors contributed equally.}
\affiliation{Universit\'e Paris-Saclay, CNRS, Laboratoire de Physique des Solides, 91405 Orsay, France}
\affiliation{The Institute for Solid State Physics, The University of Tokyo, Kashiwa, Japan}

\author{Daniel Lozano-G\'omez}
\thanks{These authors contributed equally.}
\affiliation{Institut f\"ur Theoretische Physik and W\"urzburg-Dresden Cluster of Excellence ctd.qmat, Technische Universit\"at Dresden, 01062 Dresden, Germany}

\author{Chun-Jiong Huang}
\thanks{These authors contributed equally.}
\affiliation{Department of Physics, The University of Tennessee, Knoxville, Tennessee, USA}

\author{Guanyue Chen}
\affiliation{The Institute for Solid State Physics, The University of Tokyo, Kashiwa, Japan}

\author{Hiromu Okamoto}
\affiliation{The Institute for Solid State Physics, The University of Tokyo, Kashiwa, Japan}

\author{Jian Yan}
\affiliation{The Institute for Solid State Physics, The University of Tokyo, Kashiwa, Japan}
\affiliation{Institute for Advanced Study, Chengdu University, Chengdu, China}

\author{Hikaru Takeda}
\affiliation{The Institute for Solid State Physics, The University of Tokyo, Kashiwa, Japan}

\author{Yusei Shimizu}
\affiliation{The Institute for Solid State Physics, The University of Tokyo, Kashiwa, Japan}

\author{Evan M.
Smith}
\affiliation{Department of Physics and Astronomy, McMaster University, Hamilton, Ontario L8S 4M1, Canada}
\affiliation{Brockhouse Institute for Materials Research, McMaster University, Hamilton, Ontario L8S 4M1, Canada}

\author{Avner Fitterman}
\affiliation{D\'epartement de Physique, Universit\'e de Montr\'eal, Montr\'eal, Quebec H2V 0B3, Canada}
\affiliation{Regroupement Qu\'eb\'ecois sur les Mat\'eriaux de Pointe (RQMP), Quebec H3T 3J7, Canada}

\author{Andrea D.
Bianchi}
\affiliation{D\'epartement de Physique, Universit\'e de Montr\'eal, Montr\'eal, Quebec H2V 0B3, Canada}
\affiliation{Regroupement Qu\'eb\'ecois sur les Mat\'eriaux de Pointe (RQMP), Quebec H3T 3J7, Canada}

\author{Bruce D.
Gaulin}
\affiliation{Department of Physics and Astronomy, McMaster University, Hamilton, Ontario L8S 4M1, Canada}
\affiliation{Brockhouse Institute for Materials Research, McMaster University, Hamilton, Ontario L8S 4M1, Canada}
\affiliation{Canadian Institute for Advanced Research, 661 University Avenue, Toronto, Ontario, Canada}

\author{Han Yan (闫寒)}
\email{hanyan@issp.u-tokyo.ac.jp}
\affiliation{The Institute for Solid State Physics, The University of Tokyo, Kashiwa, Japan}

\author{Minoru Yamashita}
\email{my@issp.u-tokyo.ac.jp}
\affiliation{The Institute for Solid State Physics, The University of Tokyo, Kashiwa, Japan}

\date{\today}
\begin{abstract}
We report low-temperature magnetization and magnetostriction measurements on the dipole-octupole pyrochlore $\mathrm{Ce_2Hf_2O_7}$, revealing an unconventional field response for $\mathbf B\parallel[111]$.
The magnetization shows no kagome-ice plateau; instead it evolves continuously and exhibits two rapid changes in slope near 0.35~T and 1.2~T, accompanied by magnetostriction features at the same field scales.
Classical Monte Carlo simulations, exact diagonalization, and ground-state analysis of the Hamiltonian show that a representative QSI-compatible parameter set captures the data.
For this parameter set, the lower-field anomaly marks a closely spaced transition sequence from a two-color kagome spin
liquid (KSL) through a narrow three-color KSL into a mixed two-/three-color KSL, while the upper-field anomaly marks the
transition from the mixed KSL into a nearly polarized state.
These results identify $\mathrm{Ce_2Hf_2O_7}$, and dipole-octupole pyrochlore magnets more broadly, as promising platforms for exotic KSLs beyond conventional spin ice.
\end{abstract}
\maketitle 

\end{CJK*}

\prlsection{Introduction}
Geometrically frustrated pyrochlore magnets provide a fertile setting for emergent gauge phenomena.
In spin ice, local Ising anisotropy and frustrated interactions impose the ``2-in-2-out'' rule on each tetrahedron,
producing an extensively degenerate ground-state manifold; quantum fluctuations can promote this manifold to a quantum spin
ice with fractionalized excitations and emergent electrodynamics~\cite{Ref1_Gingras2014QSI,Ref2_Castelnovo2008Monopoles,Ref3_Hermele2004PyrochlorePhotons,Ref4_Rau2019RareEarthPyrochlores}.
More recently, Ce-based dipole-octupole pyrochlores have emerged as promising candidates for QSI physics and unconventional field-induced
constrained states~\cite{Ref17_Sibille2020QuantumLiquidOctupoles,Ref18_Poree2024FractionalMatterGaugeField,
Ref19_Yahne2024DipolarSpinIceCe2Sn2O7,Ref20_Smith2022U1piCe2Zr2O7,Ref21_Bhardwaj2022SleuthingCe2Zr2O7,
Ref22_Smith2023FieldResponseCe2Zr2O7,Ref5_Huang2014DipolarOctupolar,Ref6_Benton2020GroundStateDiagram,
Ref7_Patri2020Magnetostriction,Ref8_Desrochers2022CompetingQSL,Ref9_Hosoi2022NeutronScattering,
Ref10_Sanders2023TunableQED}.

A hallmark of spin-ice physics is the response to a magnetic field along the crystallographic $[111]$ direction.
In a classical spin ice (CSI), the triangular layers are polarized while the kagome layers remain constrained by the ice rule,
producing the ``2-up-1-down'' kagome-ice state and a characteristic magnetization plateau~\cite{Moessner2003}.
The plateau reflects the persistence of the ice-rule manifold over a finite field window and has been observed in canonical
spin-ice materials $\mathrm{Dy_2Ti_2O_7}$~\cite{Matsuhira2002,Ref26_Sakakibara2003LiquidGasTransition,Tabata2006}.
It therefore provides a stringent benchmark for identifying whether a field-induced state is the conventional kagome-ice
manifold or a qualitatively different states.

\begin{figure*}[th!]
  \centering
  \includegraphics[width=\linewidth]{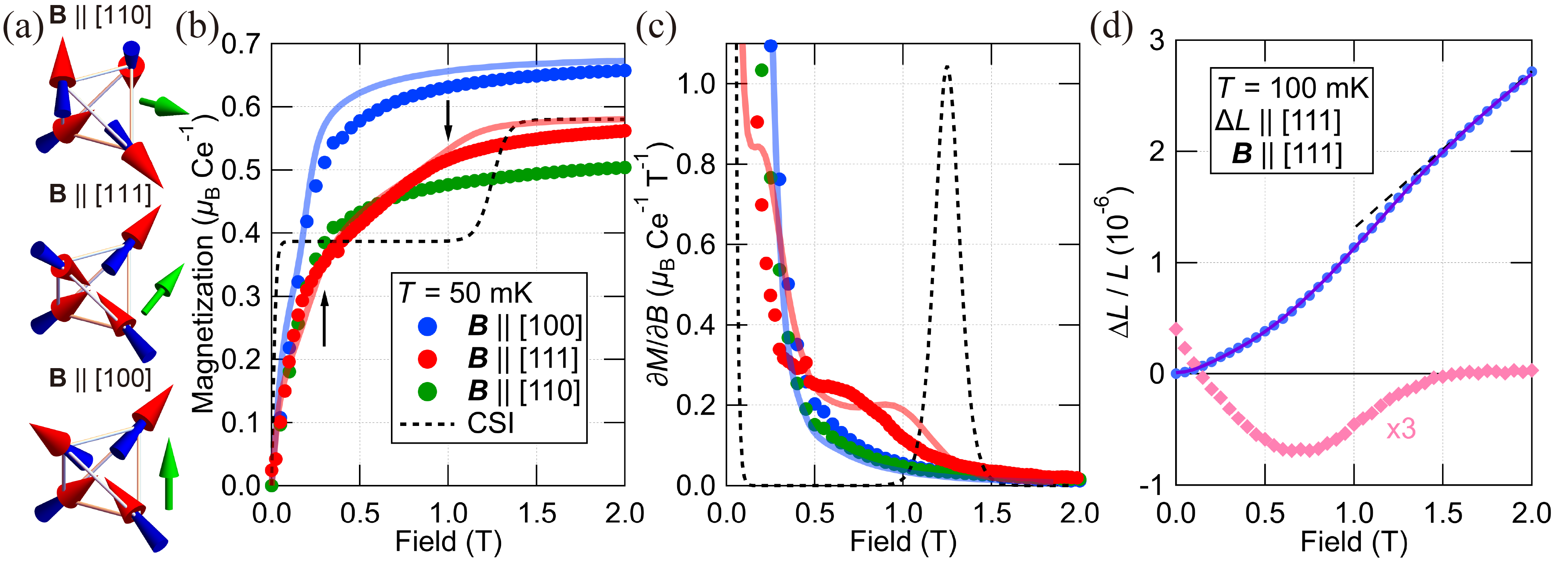}
    \caption[Unconventional field response and magnetoelastic anomalies]{
Unconventional $[111]$ field response and magnetoelastic anomalies in Ce$_2$Hf$_2$O$_7$.
(a) Field geometry and high-field spin polarization for a single tetrahedron. For $\mathbf B\parallel[110]$, the top two spins of the tetrahedron shown in the figure do not couple linearly to an ideal $[110]$ field.
(b,c) Magnetization and field derivative at 50~mK.
The data show two rapid slope-change anomalies near $\sim 0.35$~T and $\sim1$~T
for $\mathbf B\parallel[111]$; the broad shoulder between them is not treated as a separate anomaly.
Thick lines show exact-diagonalization (ED) results for
$\{J_x,J_y,J_z,J_{xz}\}=\{0.015,0.047,0.014,-0.012\}$~meV and $g_z=2.36$;
black dashed lines show the CSI result for $\mathbf B\parallel[111]$.
(d) Magnetostriction along $[111]$ under $\mathbf B\parallel[111]$ at 100~mK.
The pink diamonds show the deviation from the high-field linear fit, multiplied by 3;
the purple line is the CMC fit using the same parameter set.
}
\label{fig:1}
\end{figure*}

Here we study the dipole-octupole pyrochlore $\mathrm{Ce_2Hf_2O_7}$~\cite{Ref11_Smith2025AnnualReviewQSI,Ref12_Sibille2015Ce2Sn2O7,Ref13_Gaudet2019Ce2Zr2O7,Ref14_Gao2019ThreeDimensionalQSL,Ref15_Poree2022CrystalFieldCe2Hf2O7,Ref16_Smith2025TwoPeakHeatCapacity}
through magnetization measurements down to 50~mK and magnetostriction measurements down to 100~mK.
For $\mathbf B\parallel[111]$, the magnetization shows no plateau [Fig.~\ref{fig:1}(b)].
Instead, it evolves continuously through the field range where kagome ice would normally appear,
with two rapid slope-change anomalies near $\sim0.35$~T and $\sim1$~T.
Magnetostriction features appear at the same field scales.
These combined magnetic and magnetoelastic anomalies are the central experimental puzzle:
the $[111]$ field response shows the absence of the canonical kagome-ice plateau of a CSI and instead points to two
distinct field-induced changes of the low-temperature states.

As we demonstrate in this work, this anomalous response is rooted in the dipole-octupole character of the Ce$^{3+}$ doublet, and suggests a new class of field-induced kagome spin liquids (KSLs).
The external magnetic field couples linearly to the dipolar $S^z$ component, while dominant transverse exchange, especially in the octupolar $S^y$ channel, continues to constrain the kagome-plane spins.
The field therefore does not select a fixed kagome-ice manifold; instead, it changes the local spin-component values in the
ground state and produces a sequence of KSLs with distinct generalized ``ice rules'': a two-color KSL with a
two-red--one-blue rule on each triangle, a narrow three-color KSL with a one-red--one-blue--one-yellow rule, and a mixed
two+three-color KSL.
The two slope-change anomalies signal field-induced transition and crossover involving the KSL regimes: a lower-field
transition sequence among KSL states and an upper-field crossover out of the KSLs into a nearly polarized state.
Our results therefore identify dipole-octupole pyrochlore magnets as promising candidates for realizing these exotic KSLs.

\prlsection{Magnetization, magnetostriction, and lack of plateaus}
To probe the low-temperature field response of the dipole-octupole spins, we measured the magnetization of
single-crystalline $\mathrm{Ce_2Hf_2O_7}$ down to $T=50$~mK and the magnetostriction down to $T=100$~mK.
The magnetization measurements were performed using a Faraday-force magnetometer~\cite{shimizu_2021} under the magnetic
fields applied along the principal crystallographic directions $[100]$, $[110]$, and $[111]$ [Figs.~\ref{fig:1}(a,b,c)].
The magnetostriction along the $[111]$ direction was measured using a capacitive dilatometer~\cite{Kuchler_2012} under
$\mathbf{B}\parallel[111]$, allowing us to track the relative length change $\Delta L/L$ as a function of the magnetic field
[Fig.~\ref{fig:1}(d)].
For each measurement, the magnetization or the magnetostriction was recorded during both field-increasing and field-decreasing sweeps to investigate the magnetic hysteresis that develops below around 300~mK.
Here, we focus on the data at the lowest temperature recorded in decreasing the field because the data in increasing the field may not have reached equilibrium due to frustration effects.
Further details of sample preparation, measurement setups, and the magnetic hysteresis are provided in the Supplementary Materials (SM).

As shown in Fig.~\ref{fig:1}(b), the magnetization curves exhibit pronounced anisotropy consistent with strong local Ising
axes of the Ce moments.
For fields applied along $[111]$, however, the field dependence deviates markedly from the canonical
behavior expected for spin ice: instead of the characteristic kagome-ice magnetization plateau predicted for $[111]$ fields,
the magnetization evolves continuously and displays two rapid changes in slope near $\sim 0.35$~T and $\sim 1$~T.
The derivative $\partial M/\partial B$ highlights these two slope-change anomalies for $\mathbf{B}\parallel[111]$
[Fig.~\ref{fig:1}(c)].
Between them, the magnetization evolves through a broad intermediate-field shoulder, which is not
identified as a separate anomaly.
This directional selectivity already suggests that the observed slope changes are tied to
the special decomposition of the pyrochlore lattice into triangular and kagome planes under a $[111]$ field.

Strikingly, magnetostriction measurements reveal corresponding features at the same field scales.
The field dependence of the magnetostriction shows slope changes associated with the lower- and upper-field anomalies, which
can be more clearly seen by subtracting the linear field component measured above the saturation field (the pink diamonds in
Fig.~\ref{fig:1}(d), see SM for more details).
The coincidence of the magnetic and magnetoelastic signatures indicates a
common microscopic origin in which the field-driven changes away from the standard kagome-ice state are accompanied by a
measurable lattice response.
These experimental results therefore point not merely to a weak crossover in magnetization,
but to two distinct field scales at which the low-temperature constrained state changes.

In a CSI, a magnetic field applied along the $[111]$ direction polarizes the spins on the triangular layers while the kagome
layers remain constrained by the ice rule, producing a robust kagome-ice state and a magnetization plateau at one third of
the saturation value, as illustrated by the dashed line in Fig.~\ref{fig:1}(b)~\cite{Ref26_Sakakibara2003LiquidGasTransition}.
This plateau reflects the stability of a single $2$-in-$1$-out kagome manifold over a finite field window.
In $\mathrm{Ce_2Hf_2O_7}$, however, the same $[111]$ geometry does not select the conventional kagome-ice manifold.
The magnetization instead passes continuously through the expected plateau region and exhibits two rapid slope changes, while
the magnetostriction shows corresponding features.
This suggests that the field-induced state is different from kagome ice.
In the remaining part of this paper, we identify the lower anomaly with transitions among different field-induced KSL phases
and the upper anomaly with the crossover from the KSL phases into the nearly polarized state.\\

\prlsection{Microscopic Hamiltonian and parameter selection}
To identify the microscopic mechanisms underlying the anomalous $[111]$ field response, we consider the symmetry-allowed nearest-neighbor Hamiltonian for dipole-octupole doublets on the pyrochlore lattice,
\[
\begin{split}
    \mathcal H =
\sum_{\langle ij\rangle}
(&
J_{x}S_i^xS_j^x +
J_{y}S_i^yS_j^y +
J_{z}S_i^zS_j^z \\
&+
J_{xz}(S_i^xS_j^z+S_i^zS_j^x)
)
-
\mu_B g_z\sum_i (\mathbf B\cdot \hat z_i) S_i^z .
\end{split}
\]
Here $S_i^{x,y,z}$ are effective pseudospin-$1/2$ components, $\hat z_i$ is the local $[111]$ axis, $\mu_B$ is the Bohr magneton, and $g_z=2.36$.
In this local basis, $S^z$ is the magnetic dipole component that couples linearly to the field, while $S^y$ has octupolar character~\cite{Ref5_Huang2014DipolarOctupolar}.
For the classical simulations and ground-state analysis, the pseudospins are treated as classical vectors of length $S=1/2$, using the normalization $M_\parallel=g_zS\mu_B$.

\begin{figure}[th!]
    \centering
    \includegraphics[width=\columnwidth]{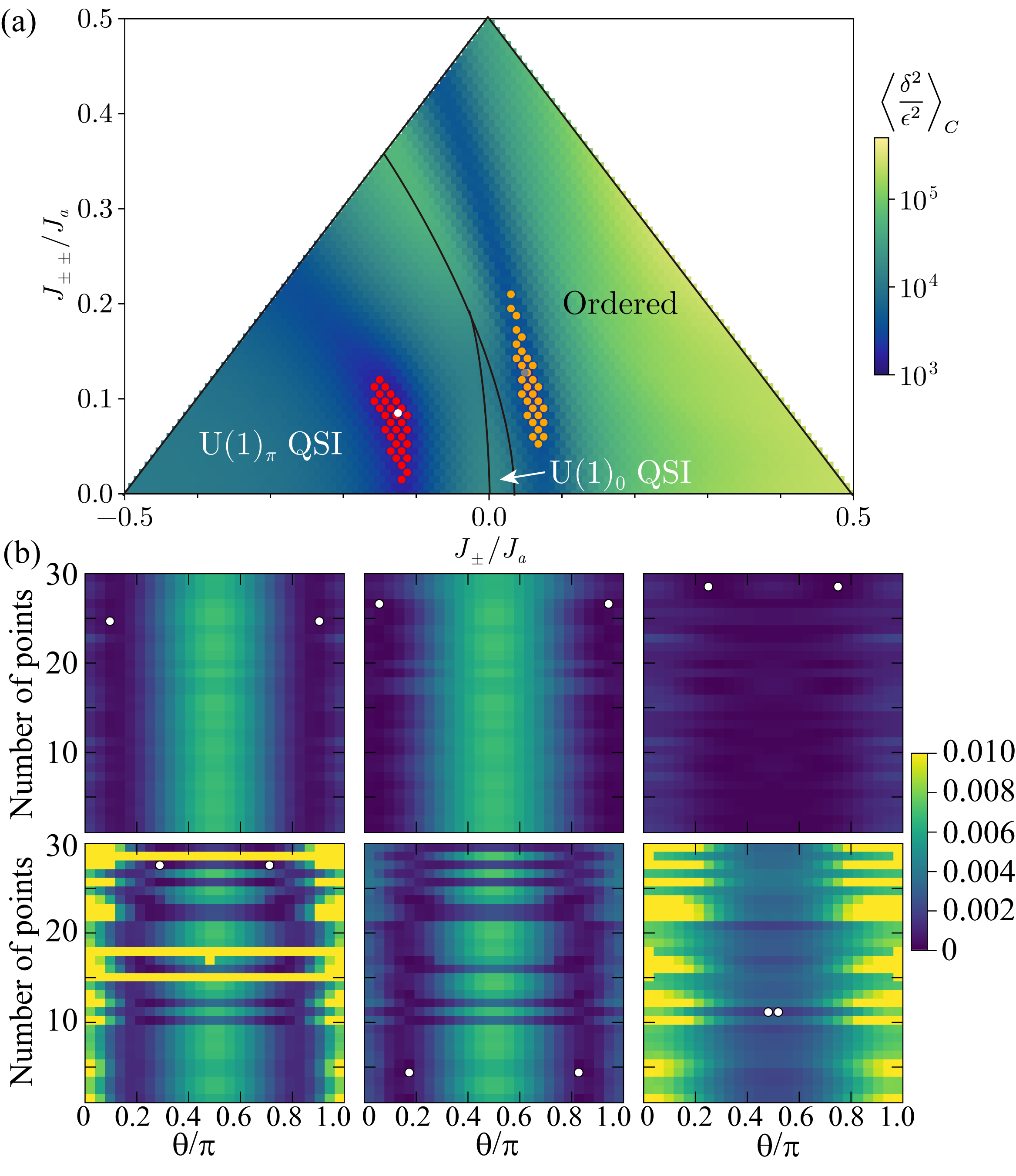}
    \caption[Hamiltonian parameter fitting from field-dependent magnetization]{\label{fig:CMC_fit}
    (a) Heat-capacity goodness-of-fit landscape from Ref.~\cite{Ref16_Smith2025TwoPeakHeatCapacity}.
Red and orange points are the parameter sets sampled around the best QSI and ordered points, respectively.
$J_a$ is the dominant exchange and $J_\pm$, $J_{\pm\pm}$ are transverse exchanges in this basis (see SM for definitions).
    (b) $\theta$  dependence of the magnetization misfit $\chi^2$ for the QSI-region points (top) and ordered-region points (bottom) for the three non-equivalent permutations
    $\{J_{\tilde{x}}, J_{\tilde{y}}, J_{\tilde{z}}\} = \{J_a, J_b, J_c\}$ (left), $\{J_a, J_c, J_b\}$ (middle), and $\{J_b, J_a, J_c\}$ (right).
    $\theta$ is defined by $\tan 2\theta = 2J_{xz}/(J_x- J_z)$, which is undetermined by previous thermodynamics  measurements (see SM for details). 
White circles mark the minima.
    }
\end{figure}

Earlier heat-capacity fits narrowed the exchange parameters of $\mathrm{Ce_2Hf_2O_7}$ to QSI-compatible and nearby ordered
regions~\cite{Ref16_Smith2025TwoPeakHeatCapacity}.
These fits constrain the overall exchange scale and parameter space but
leave the exchange-axis permutation and the local $x$-$z$ rotation angle under-determined. 
The field-dependent magnetization
provides an additional constraint because it is directly sensitive to the orientation of the dipolar component relative to
the exchange eigen-basis; details of the parameterization are given in the SM.

We therefore performed classical Monte Carlo (CMC) simulations over the parameter points in Fig.~\ref{fig:CMC_fit}(a),
scanning over all the allowed permutations and rotation angles.
The misfit $\chi^2$ was evaluated using the field-decreasing
magnetization data for $\mathbf B\parallel[100]$ and $[111]$; the $[110]$ data were excluded because this direction is
strongly affected by small misalignment, as discussed in the SM.
The QSI-region parameters reproduce the experimental data
better than the ordered-region parameters throughout the scan [Fig.~\ref{fig:CMC_fit}(b)].
The representative best-fit parameter set used below is
$\{J_x,J_y,J_z,J_{xz}\}=\{0.015,0.047,0.014,-0.012\}$~meV, close to estimates from other
works~\cite{Ref23_Poree2023ExchangeHierarchyCe2Hf2O7,Ref24_Bhardwaj2025ThermodynamicsCe2Hf2O7}.
Exact-diagonalization (ED) calculations with this set reproduce the main magnetization features, including the two $[111]$
slope-change anomalies [Fig.~\ref{fig:1}(b,c)], motivating the KSL analysis below.\\

\begin{figure}[th!]
    \centering
    \includegraphics[width=\columnwidth]{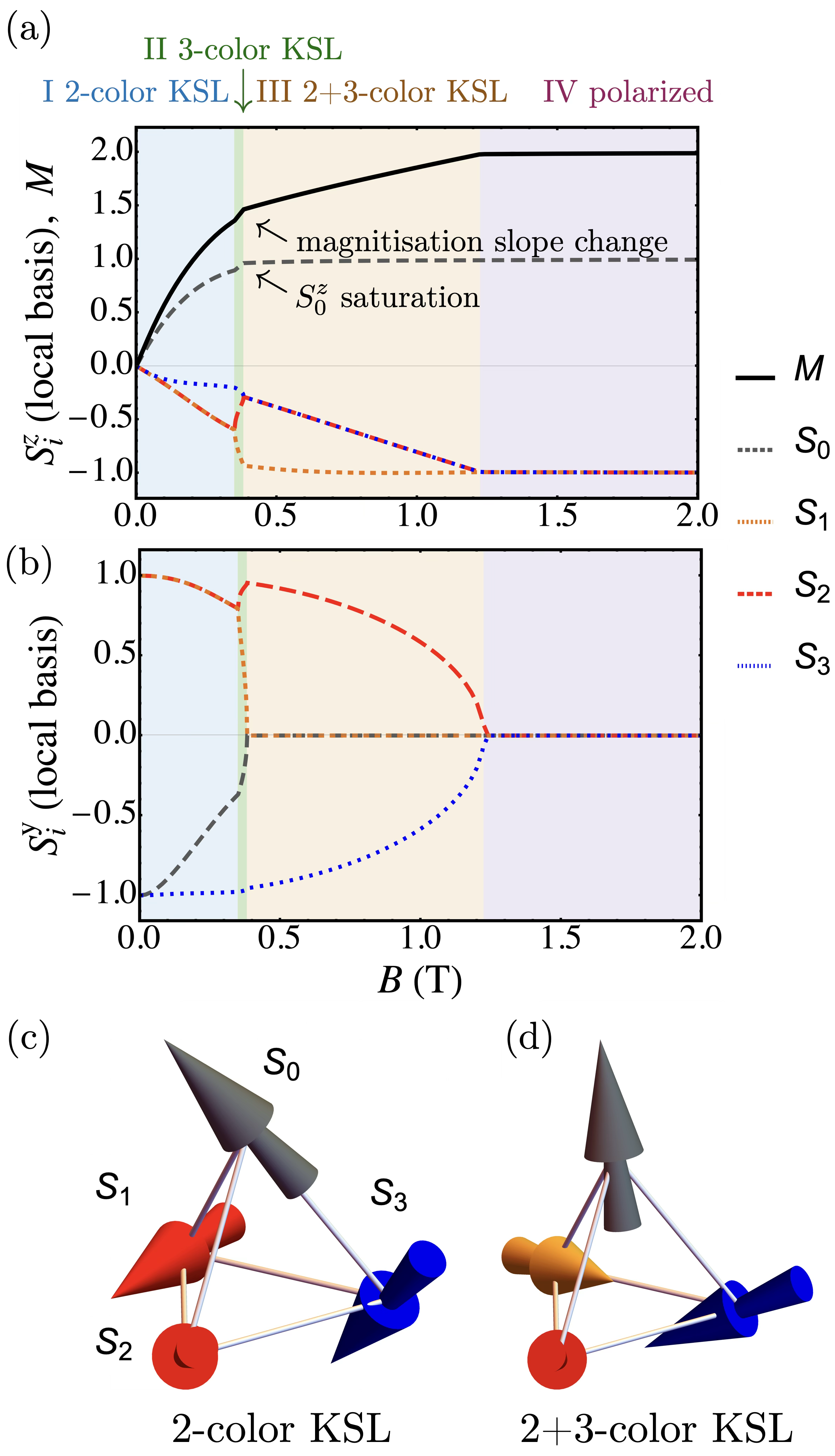}
    \caption[Zero-temperature color-constrained kagome spin liquid states]{\label{fig:theory_frag}
    Zero-temperature color-constrained kagome spin liquid states from the classical ground-state analysis.
    (a,b) Field evolution of the ground-state $S^z_\mu$ and $S^y_\mu$ components; sublattice $0$ is the triangular layer and sublattices $1$--$3$ are kagome-layer sites.
    Phase I is a two-color kagome spin liquid, phase II is a three-color kagome spin liquid, and phase III is a mixed two-/three-color kagome spin liquid with two-color $S^z$ and three-color $S^y$ local constraints.
    (c,d) Representative real-space configurations at $B=0.25$~T and $0.75$~T.
    }
\end{figure}

\begin{figure*}[ht!]
    \centering\begin{overpic}[width=\textwidth]{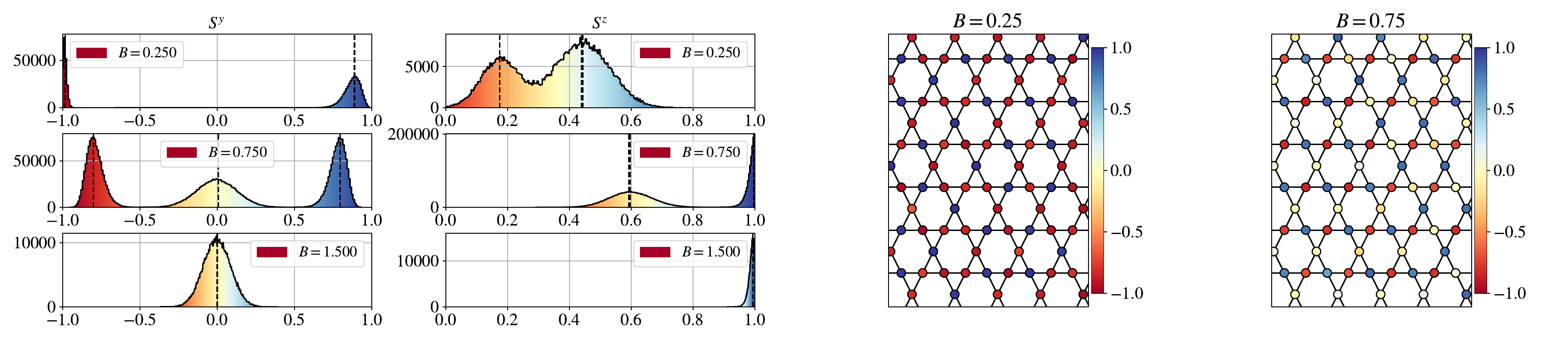}
    \put(4,21){(a)}
    \put(27,21){(b)}
    \put(54,21){(c)}
    \put(78,21){(d)}
    \put(58,0){2-color KSL}
    \put(82,0){2+3-color KSL}
    \end{overpic}
    \caption[Classical Monte Carlo evidence for multi-color kagome spin liquids]{\label{fig:CMC_statistics}
    Classical Monte Carlo evidence for the two-color and mixed two-/three-color kagome spin liquids.
    (a,b) Low-temperature distributions of $S^y$ and $S^z$ on a kagome plane at representative fields; dashed lines mark the $T=0$ single-tetrahedron values.
    (c,d) Snapshots of the corresponding $S^y$ configurations.
The low-field two-color rule is replaced at intermediate field by a one-red--one-blue--one-yellow rule in the octupolar sector.
    }
\end{figure*}

\prlsection{Field-induced multi-color KSLs}
With the representative Hamiltonian specified, we now show that the anomalous $[111]$ response reflects a sequence of KSL phases with distinct generalized ``ice rules''.
The zero-temperature classical ground state shows four field regimes, labeled phases I--IV in Fig.~\ref{fig:theory_frag}(a,b).
The lower-field anomaly corresponds to a closely spaced I$\rightarrow$II$\rightarrow$III transition sequence near
0.35--0.38~T, while the upper-field anomaly corresponds to the III$\rightarrow$IV crossover near 1.2~T.
In the finite-temperature simulations and experiment, these transition and crossover are rounded but remain visible as two rapid
slope-change anomalies in the $[111]$ magnetization.

\textit{Phase I: two-color KSL.
}
In the low-field regime, up to~$\sim0.35$~T in the classical ground-state analysis, the kagome-plane spins form a
``two-color KSL'': on each kagome triangle, two sites take the same spin configuration in local basis while the third takes
a different one, giving the two-red--one-blue rule shown in Fig.~\ref{fig:theory_frag}(c).
In Fig.~\ref{fig:theory_frag}(a,b), this appears as a degeneracy between the red and yellow branches, corresponding to
sublattices 1 and 2 (two-red in the rule), while the blue branch, corresponding to sublattice 3, is distinct (one-blue in the rule).
Here, the labeling is artificial, as permuting the three spin configuration and also the global symmetry action $S^y \rightarrow - S^y$ does not change the energy of the kagome triangle, leading to 6 ground states on a  kagome triangle. 
This rule constrains both the dipolar $S^z$ and octupolar $S^y$ components, even though $S^z$ is already partly
field-polarized.
At the $T=0$ classical level, this phase also selects one of two configurations related by the global
$\mathbb{Z}_2$ transformation $S_i^y\rightarrow -S_i^y$.
Within this phase, the allowed ground-state spin-component values
evolve continuously with field, so the magnetization changes smoothly rather than forming a fixed kagome-ice plateau.

\textit{Phase II: three-color KSL.
}
Near $B\simeq0.35$--$0.38$~T, the system enters phase II, a narrow ``three-color KSL''.
In this regime the three kagome-plane spins on each triangle take three distinct local spin-component values, so each kagome triangle obeys a one-red--one-blue--one-yellow rule.
In Fig.~\ref{fig:theory_frag}(a,b), the onset of phase II appears as the splitting of red and yellow curves that were overlapping in phase I.

This narrow regime occurs when the triangular-layer dipolar component is nearly saturated, so further magnetization leads to
a transition to a different kagome-plane ground-state manifold.
The closely spaced I$\rightarrow$II and II$\rightarrow$III
transitions provide the microscopic origin of the lower-field slope-change anomaly near 0.35~T.

\textit{Phase III: mixed two-/three-color KSL.
}
For $B\simeq0.38$--$1.2$~T, the kagome plane combines the two constraints: the dipolar $S^z$ components form a two-color KSL,
while the octupolar $S^y$ components retain the three-color pattern [Fig.~\ref{fig:theory_frag}(a,b)].
The broad intermediate-field shoulder in the magnetization lies within this extended mixed two-/three-color KSL regime and is
not a separate phase boundary.
Near $\sim1.2$~T, the kagome-plane $S^z$ and $S^y$ branches merge and the system enters the
nearly polarized phase IV, in which the kagome plane no longer hosts a KSL.
This III--IV transition provides the microscopic origin of the upper-field slope-change anomaly.

The color rules also determine the emergent gauge structures.
Phase I maps to the two-up--one-down kagome-ice manifold,
although its microscopic spin configurations differ from those of conventional kagome ice, and carries the corresponding
emergent $U(1)$ gauge structure.
Phase II is more exotic: the one-red--one-blue--one-yellow rule is equivalent to two
independent local color-charge conservation laws and therefore realizes two independent copies of an emergent $U(1)$ gauge
structure, analogous to bionic Coulomb phases in three-dimensional pyrochlore four-color spin-ice
models~\cite{Khemani_2012bionic,Lozano-Gomez2024,potts2026nstatepottsicesgeneralizations}.
Phase III combines a two-color
$S^z$ constraint with a three-color, two-copy $U(1)$ constraint in the octupolar $S^y$ sector.

Full-lattice CMC simulations support this picture (Fig.~\ref{fig:CMC_statistics}).
In phase I, the kagome-plane $S^y$ and $S^z$ distributions are bimodal, consistent with the two-color rule.
In phase III, the $S^y$ distribution becomes trimodal while $S^z$ remains bimodal, the characteristic fingerprint of the mixed two-/three-color KSL.
The snapshots show the same transition from a two-color constraint to a one-red--one-blue--one-yellow rule.

\prlsection{Discussion}
Our results show that $\mathrm{Ce_2Hf_2O_7}$ departs sharply from the canonical $\mathbf B\parallel[111]$ response of classical spin ice.
Instead of a kagome-ice plateau, the magnetization evolves continuously and exhibits two rapid slope-change anomalies near $\sim 0.35$~T and $\sim 1$~T, accompanied by magnetostriction features at the same field scales.
In the representative dipole-octupole Hamiltonian, these anomalies correspond to phase transitions involving the various
exotic KSL regimes: a lower-field transition sequence from the two-color KSL through the narrow three-color KSL into the
mixed regime, and an upper-field transition from the mixed two-/three-color KSL into the nearly polarized phase.
Future neutron-scattering measurements could test the dipolar correlations of these KSLs, while octupolar three-color correlations may require complementary magnetoelastic or thermodynamic probes.
These results suggest dipole-octupole pyrochlores as promising platforms for exotic field-induced KSLs beyond the conventional spin-ice paradigm.\\

\begin{acknowledgments}
We thank Cristian D.
Batista and Y.
B.
Kim for fruitful discussions.
This work was supported by Grants-in-Aid for Scientific Research (KAKENHI) (numbers JP23K25813 and JP24K06934) and the Natural Sciences and Engineering Research Council of Canada.
E.K. acknowledges support from the Japan Society for the Promotion of Science (JSPS) for Foreign Researchers.
D.L.-G. acknowledges financial support from the DFG through the W\"urzburg-Dresden Cluster of Excellence on Complexity,
Topology and Dynamics in Quantum Matter -- \textit{ctd.qmat} (EXC 2147, project-id 390858490) and
through SFB 1143 (project-id 247310070).
D.L.-G. is supported by the Hallwachs-R\"ontgen Postdoc Program of ct.qmat.
C.J.H. was supported by the U.S. Department of Energy, Office of Science, Office of Basic Energy Sciences, under Award Number DE-SC0022311.
H.Y. acknowledges the 2024 Toyota Riken Scholar Program from the Toyota Physical and Chemical Research Institute, and the Grant-in-Aid for Research Activity Start-up from Japan Society for the Promotion of Science (Grant No. 24K22856).
\end{acknowledgments}

\bibliography{CHO_ref}

\onecolumngrid
\clearpage
\begin{center}
	\Large{\textbf{End Matter for ``\thistitle''}}
\end{center}
\setcounter{equation}{0}
\setcounter{figure}{0}
\setcounter{table}{0}
\makeatletter
\renewcommand{\theequation}{E\arabic{equation}}
\renewcommand{\thefigure}{E\arabic{figure}}
\renewcommand{\thetable}{E\arabic{table}}
\renewcommand{\bibnumfmt}[1]{[#1]}
\renewcommand{\citenumfont}[1]{#1}

Here, we will discuss the theoretical understanding of the magnetostriction data. 
Our theory follows closely Ref.~\cite{Ref7_Patri2020Magnetostriction}: considering the symmetry of the
model, one can write down the most general terms for the dipolar and
octupolar moments coupling to the elastic strain.
Then, given the
average magnetic moment on each sublattice site at certain external magnetic field, the elastic energy is minimized at a finite
strain, which produces the macroscopic lattice distortion.
More specifically, the lattice distortion measured along $[111]$ under the
magnetic field applied along $[111]$ is given by the expression below.
\[
\begin{split}
\left( \frac{\Delta L}{L} \right) =& \frac{g_{z}B_{z}}{27c_{B}}\left\lbrack
\left( g_{10} + 2g_{9} \right)\left( 3\tau_{(0)}^{z} - \tau_{(1)}^{z} - \tau_{(2)}^{z} - \tau_{(3)}^{z} \right)
+ \left( g_{4} + 2g_{3} \right)\left( 3\tau_{(0)}^{x} - \tau_{(1)}^{x} - \tau_{(2)}^{x} - \tau_{(3)}^{x} \right)
\right\rbrack\\
&+ \frac{4}{27c_{44}}g_{z}B_{z}\lbrack\left( 8\sqrt{2}g_{1} - 4g_{2} \right)\left( \tau_{(1)}^{x} + \tau_{(2)}^{x} + \tau_{(3)}^{x} \right) + \left( g_{4} - g_{3} \right)\left( 9\tau_{(0)}^{x} + \tau_{(1)}^{x} + \tau_{(2)}^{x} + \tau_{(3)}^{x} \right)\\
&+ \left( {8\sqrt{2}g}_{7} - 4g_{8} \right)\left( \tau_{(1)}^{z} + \tau_{(2)}^{z} + \tau_{(3)}^{z} \right) + \left( g_{10} - g_{9} \right)\left( 9\tau_{(0)}^{z} + \tau_{(1)}^{z} + \tau_{(2)}^{z} + \tau_{(3)}^{z} \right)\rbrack,
\end{split}
\]
where \(\tau_{(i)}^{\alpha} = \langle S_{i}^{\alpha} \rangle \) is the averaged expectation value of
\(S_{i}^{\alpha}\), \(g_{i}\) (\(i = 1,\ldots,10\)) are magnetoelastic
coupling constants, and   \(c_{B}\) and \(c_{44}\) are components of the elastic modulus tensor.
The field-energy conversion factor \(\mu_B\) is absorbed into the phenomenological magnetoelastic coefficients. \(B_z\) denotes the corresponding local-field projection.

\begin{figure}[th!]
    \centering
    \subfloat{\includegraphics[width=0.33\columnwidth]{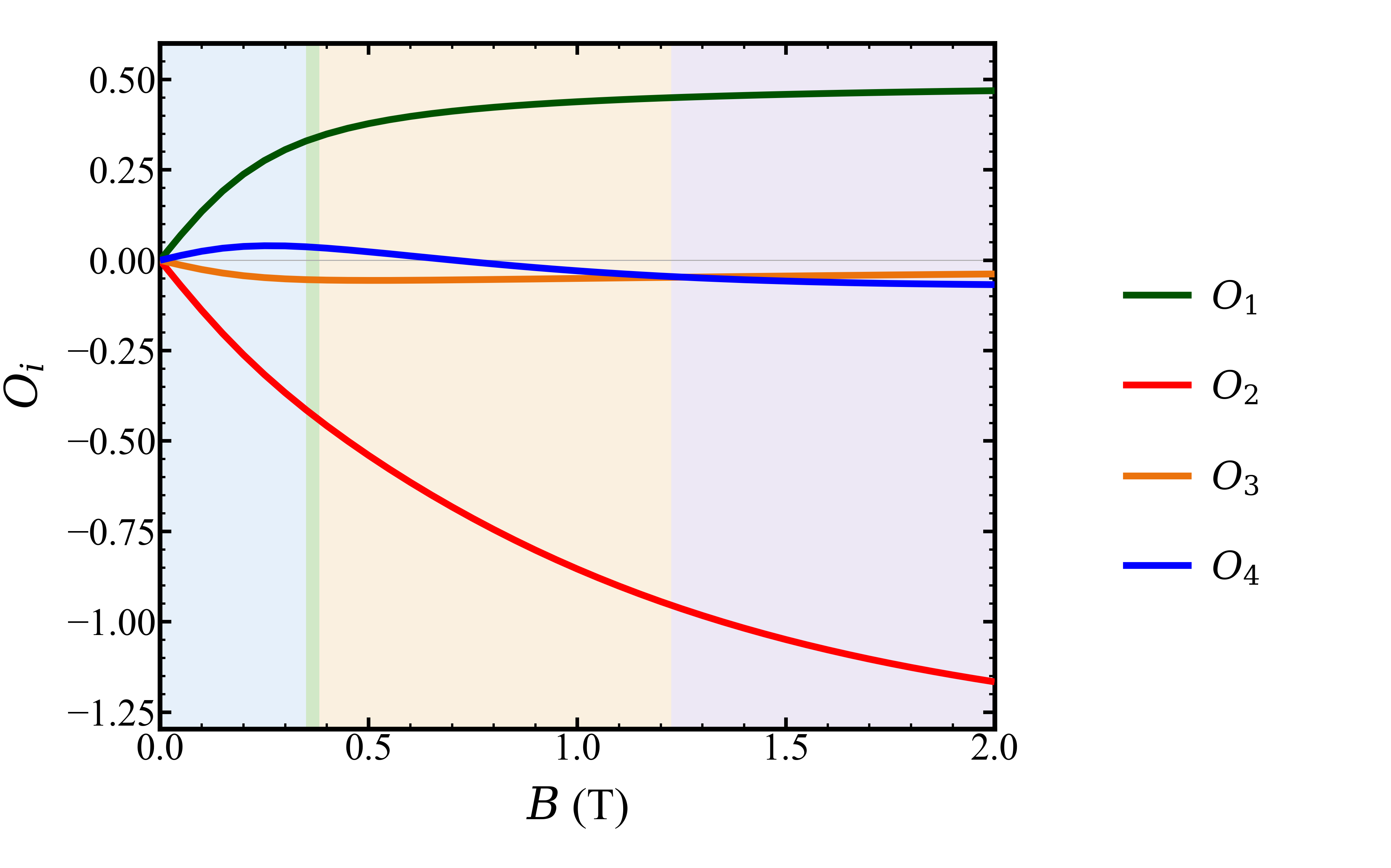}} 
    \subfloat{\includegraphics[width=0.33\columnwidth]{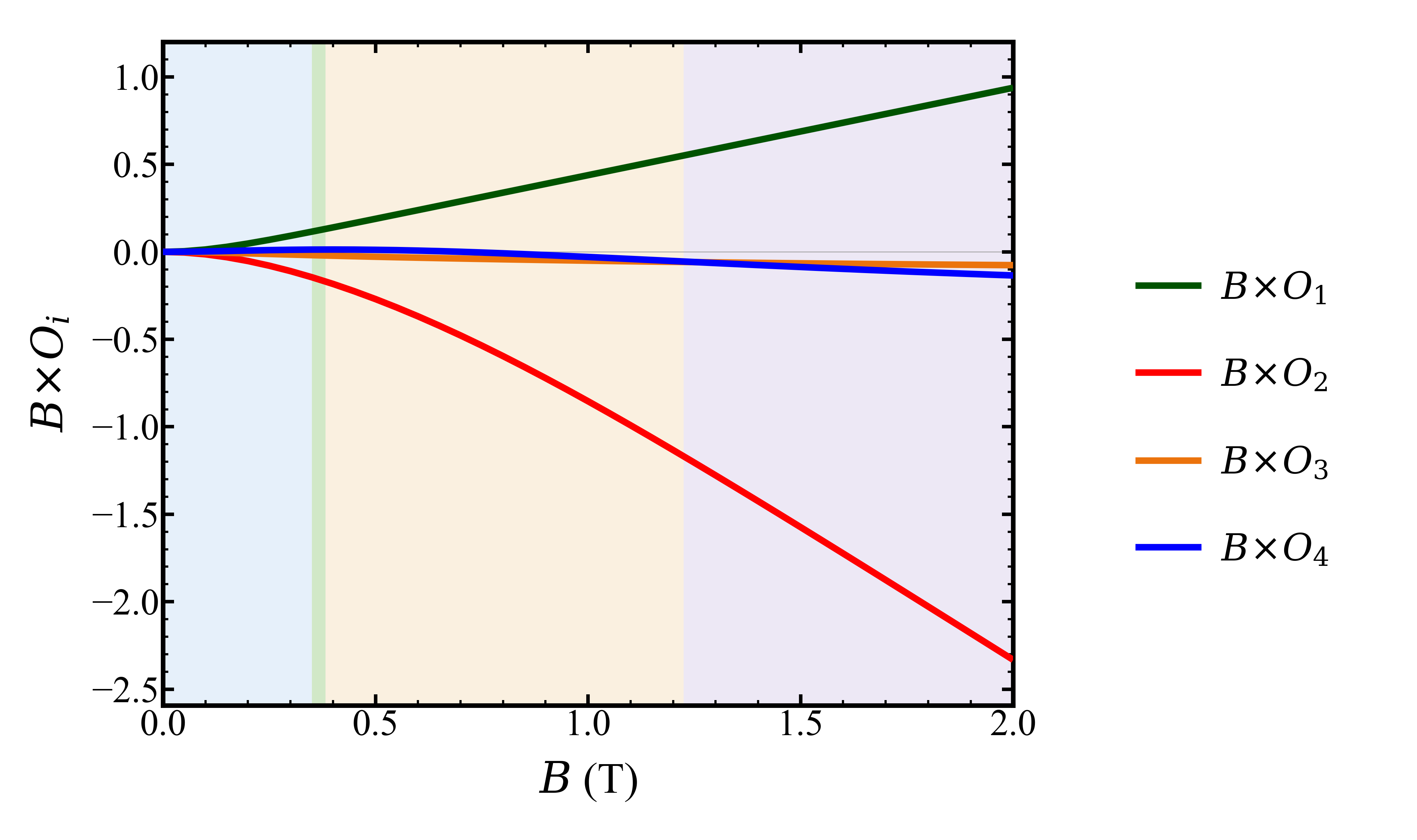}} 
    \subfloat{\includegraphics[width=0.33\columnwidth]{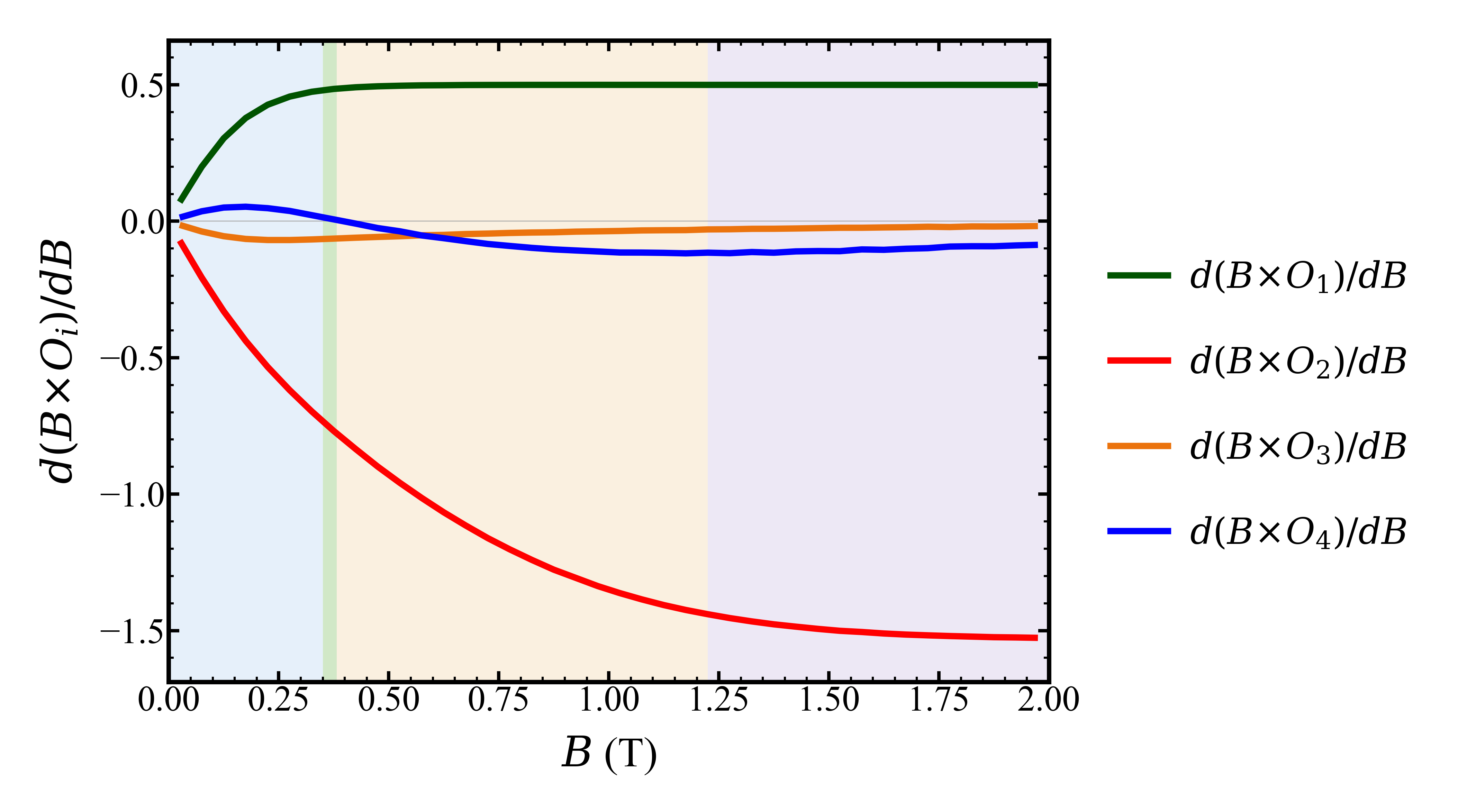}} 
    \caption{\label{fig:MCBO}
     $O_i$, $B\times O_i$ and $d(B\times O_i)/dB$ controlling the magnetostriction and their derivative of $B$. The data is obtained from Monte Carlo simulation at 100 mK.}
\end{figure}

This longitudinal response may be further written compactly as
\[
\left( \frac{\Delta L}{L} \right)_{[111]}
= g_z B \sum_{i=1}^{4} \lambda_i O_i ,
\]
where the phenomenological coefficients $\lambda_i$ absorb the elastic constants and magnetoelastic coupling constants.
The relevant spin observables are $O_1=\tau_{(0)}^z$, $O_2=\tau_{(1)}^z+\tau_{(2)}^z+\tau_{(3)}^z$, $O_3=\tau_{(0)}^x$, and
$O_4=\tau_{(1)}^x+\tau_{(2)}^x+\tau_{(3)}^x$.
Thus the magnetic contribution to $\Delta L/L$ can be described by a linear combination of the calculated curves $B\times O_i$ in Fig.~\ref{fig:MCBO}.

Given a temperature and external field, \(O_i\)'s can be
obtained from CMC simulations performed with the representative spin interaction parameters obtained in the main text.
The remaining unknowns are the grouped coupling constants $\lambda_i$, which we determine phenomenologically by fitting the experimental data. The best fit yield 
\begin{equation} 
    \lambda_1 = -3.567\times 10^{-6},\qquad
    \lambda_2 = -1.148\times 10^{-6},\qquad
    \lambda_3 = -1.968\times 10^{-5},\qquad
    \lambda_4 = -2.344\times 10^{-6}.
\end{equation}
The theory reproduced magnetostriction data is shown in Fig.~\ref{fig:1}(d).

The same set of spin exchange parameters can reproduce the magnetization and magnetostriction data satisfactorily. 
The magnetostriction experiments are very useful, as they provide a probe into order parameters $O_i$ [Fig.~\ref{fig:MCBO}], different from the magnetization measurement, which essentially measures $O_1 + O_3/3$. 
The evolution of these order parameters is a direct test of the different multiple-color KSL thesis presented in Fig.~\ref{fig:theory_frag}.

However, due to the magnetostriction model involves new unknown parameters, our current one direction measurement serves as a necessary condition check of our KSL theories. 
To further test the theory, more experiments done in other directions of distortion and external field are necessary.

\clearpage
\begin{center}
	\Large{\textbf{Supplementary Materials for ``\thistitle''}}
\end{center}
\setcounter{equation}{0}
\setcounter{figure}{0}
\setcounter{table}{0}
\makeatletter
\renewcommand{\theequation}{S\arabic{equation}}
\renewcommand{\thefigure}{S\arabic{figure}}
\renewcommand{\thetable}{S\arabic{table}}
\renewcommand{\bibnumfmt}[1]{[#1]}
\renewcommand{\citenumfont}[1]{#1}

These Supplementary Materials include
\begin{description}
    \item[S1] Materials and Methods
    \item[S2] Magnetization data at higher temperatures
    \item[S3] Effect of relaxation time on the magnetization measurements
    \item[S4] Reproducibility of the magnetization data
    \item[S5] Additional magnetostriction data
    \item[S6] Classical Monte Carlo and exact diagonalization calculations
\end{description}
Unless otherwise specified, all field-dependent low-temperature magnetization data discussed below were taken in the field-decreasing process.

\section{S1: Materials and Methods}
Single crystals of Ce$_2$Hf$_2$O$_7$ were grown by the optical floating zone method from polycrystalline samples as described
in Ref.\,\cite{Ref16_Smith2025TwoPeakHeatCapacity}.
All our samples are transparent and yellow-green in color.
Sample A (B) was cut into a rectangular plate to have a large surface perpendicular to the [110] ([111]) axis, with negligible
demagnetization effect.

The field dependence of the magnetization ($M(B)$) was measured in both samples using a commercial SQUID magnetometer above
2~K and a home-built Faraday-force magnetometer\,\cite{shimizu_2021} below 2~K.
The magnetization measured by the Faraday-force magnetometer was calibrated with the data obtained by SQUID at the same
temperature of 2~K.
At each temperature, the field dependence of the magnetization was measured in the field-increasing
process after zero-field cooling and then in the field-decreasing process.
We have confirmed that the magnetization curve is
virtually the same for the data taken after the zero-field cooling and finite field cooling, showing the absence of a glass
state.
All magnetization data presented in the main text were obtained from measurements of sample A.
The reproducibility of
the magnetization was confirmed by measurements on sample B (section S4).

The magnetostriction was measured in sample B along the [111] axis under $\mathbf{B} \parallel [111]$ using a commercial
dilatometer\,\cite{Kuchler_2012}.
The distortion of the sample along the [111] axis normalized by its length ($\Delta L/L$)
was measured as a function of the magnetic field applied parallel to the distortion.
At each temperature, $\Delta L/L$ was
also measured in the field-increasing process after zero-field cooling and then in the field-decreasing process.
We again
have confirmed that there is no difference between the magnetostriction measured after the zero-field cooling and finite field
cooling.
The magnetostriction data above 300~mK were taken by continuously sweeping the magnetic field, whereas the data at
100~mK and 200~mK were taken at fixed magnetic fields.

\section{S2: Magnetization data at higher temperatures}
\begin{figure}[th!]
          \centering
     \includegraphics[width=0.8\columnwidth]{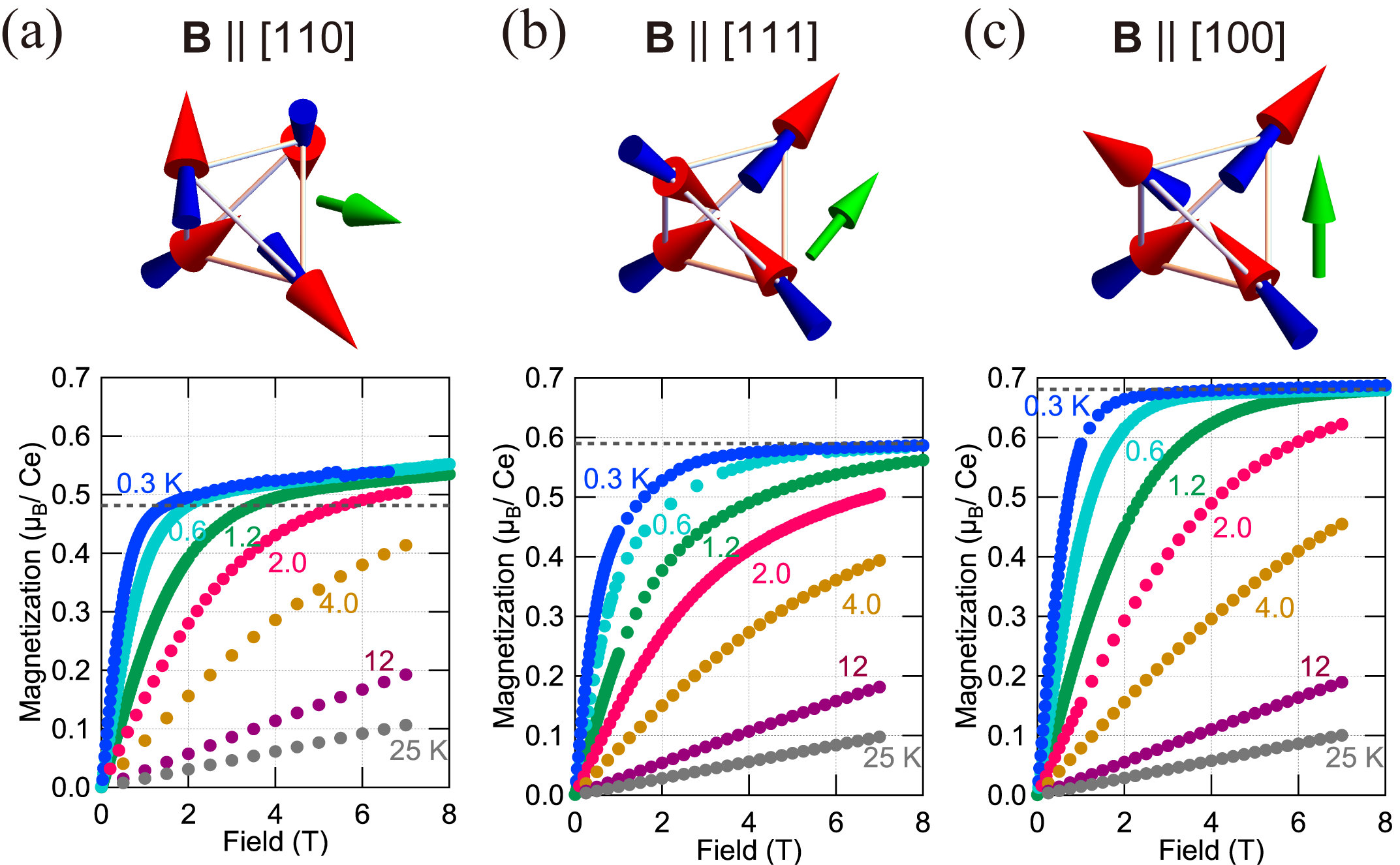}
     \hfill
         \caption{\label{fig:MH_highT}
    The field dependence of the magnetization for 0.3--25~K with the magnetic field applied along the [110] (a), [111] (b),
    and [100] (c) axes.
The data obtained in the field-increasing process are shown.
The top figures illustrate the angular
    relation between the magnetic field direction and the saturated spins in high fields for a single tetrahedron.
    The transparent arrows in (a) show the spins that do not couple to the external magnetic field in the ideal [110]
    direction, so their final saturated direction depends on the misalignment of the applied external field from the [110]
    direction.
    }
\end{figure}

Figure~\ref{fig:MH_highT} shows the field dependence of the magnetization, $M(B)$, measured for 0.3--25~K with the magnetic field applied along the three principal axes of the pyrochlore lattice.
As shown in Fig.~\ref{fig:MH_highT}, $M(B)$ increases more rapidly at lower temperatures, and then saturates above around 3~T at 300~mK.
The saturated moment depends on the magnetic field direction, reflecting the local Ising anisotropy.
As studied in classical spin ice (CSI) compounds\,\cite{Ref25_Fukazawa2002Dy2Ti2O7Anisotropy,Ref26_Sakakibara2003LiquidGasTransition},
this Ising anisotropy sets the magnitude of the saturated moment as $0.408 M_\parallel$, $0.5 M_\parallel$,
$0.577 M_\parallel$ for the magnetic field applied along [110], [111], [100] axes, respectively (top panels in
Fig.~\ref{fig:MH_highT}), where $M_\parallel = g_z S \mu_\textrm{B}$ is the total magnetic moment of each Ising spin.
These saturated moments for $M_\parallel = 1.18 \mu_\textrm{B}$\,\cite{Ref15_Poree2022CrystalFieldCe2Hf2O7}
(the dashed lines in Fig.~\ref{fig:1}) are close to the expected values, except the magnetization measured under
$\mathbf{B} \parallel [110]$, which is known to be very sensitive to minute misalignment of the field direction from the
[110] axis\,\cite{Ref25_Fukazawa2002Dy2Ti2O7Anisotropy,Ref24_Bhardwaj2025ThermodynamicsCe2Hf2O7,Ref27_Placke2020HierarchyEnergyScales}.
This good agreement between the anisotropic magnetization data and the expected values validates the Ising anisotropy of the magnetic moment in this material.

\begin{figure}[th!]
          \centering
     \includegraphics[width=0.4\columnwidth]{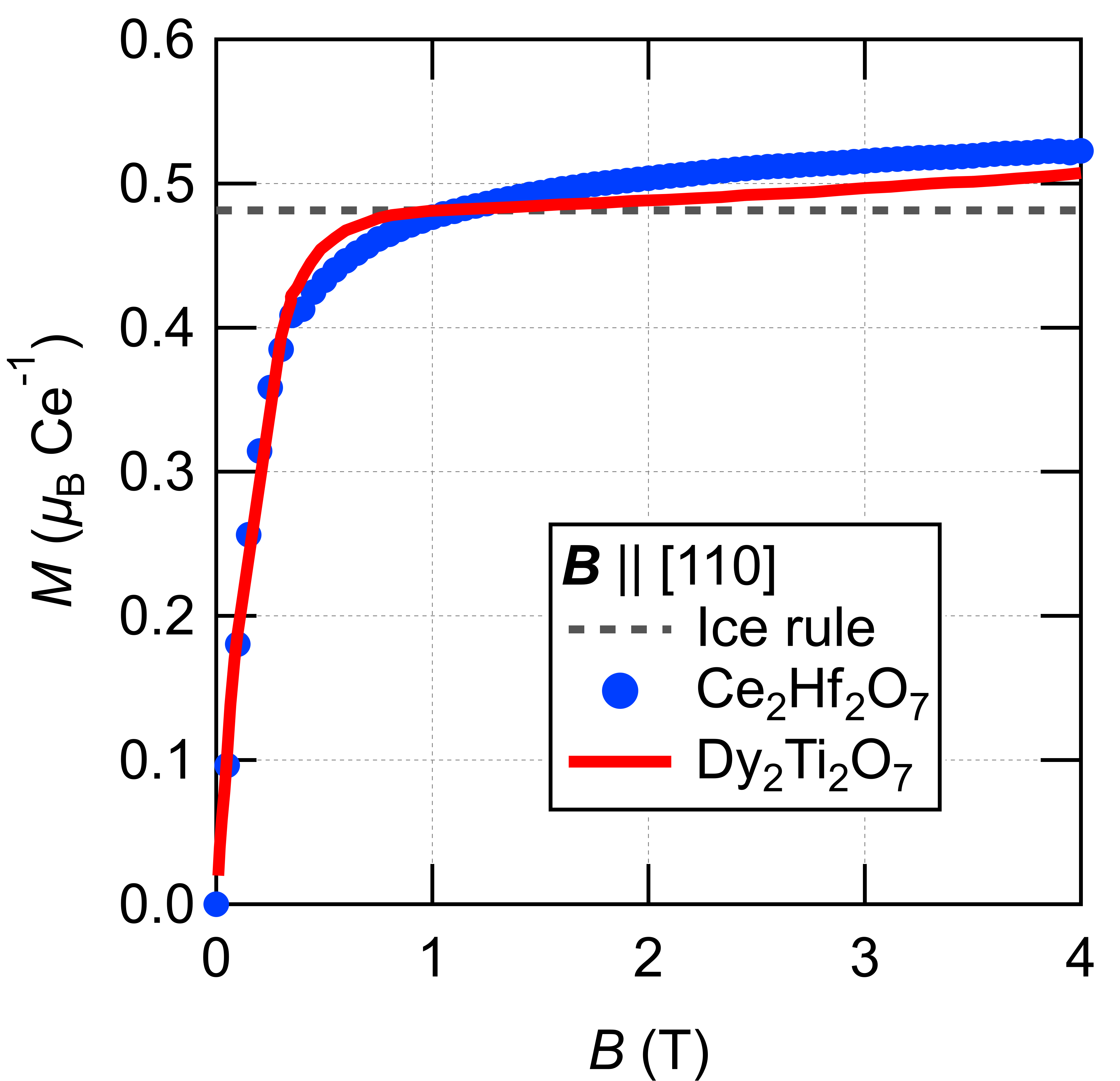}
     \hfill
         \caption{\label{fig:MH_misalign}
    The field dependence of the magnetization under $\mathbf{B} \parallel [110]$.
    The blue circles show the magnetization data of Ce$_2$Hf$_2$O$_7$ at 50~mK obtained during the field-decreasing process.
    The grey dashed line shows the saturation value expected by the ice rule.
The red line shows the magnetization data of
    Dy$_2$Ti$_2$O$_7$ at 2~K, taken from Ref.~\cite{Ref25_Fukazawa2002Dy2Ti2O7Anisotropy}, measured with $0.5^\circ$
    misaligned from [110] toward [100].
The vertical axis of the Dy$_2$Ti$_2$O$_7$ data is rescaled so that the saturation
    moment of Dy$_2$Ti$_2$O$_7$ ($4.08~\mu_\textrm{B}$ Dy$^{-1}$) matches the value for Ce$_2$Hf$_2$O$_7$
    ($0.481~\mu_\textrm{B}$ Ce$^{-1}$).
    }
\end{figure}

To estimate the angle deviation of our magnetization measurements under $\mathbf{B} \parallel [110]$, we compare our data
with that obtained in the classical spin ice compound Dy$_2$Ti$_2$O$_7$ with the intentional misalignment of $0.5^\circ$
from [110]~\cite{Ref25_Fukazawa2002Dy2Ti2O7Anisotropy}.
As shown in Fig.~\ref{fig:MH_misalign}, the rescaled Dy$_2$Ti$_2$O$_7$ curve (red line) reproduces the excess magnetization in our data well.
We therefore estimate the misalignment to be about \(1^\circ\).

\begin{figure}[bht!]
    \centering
    \includegraphics[width=0.8\columnwidth]{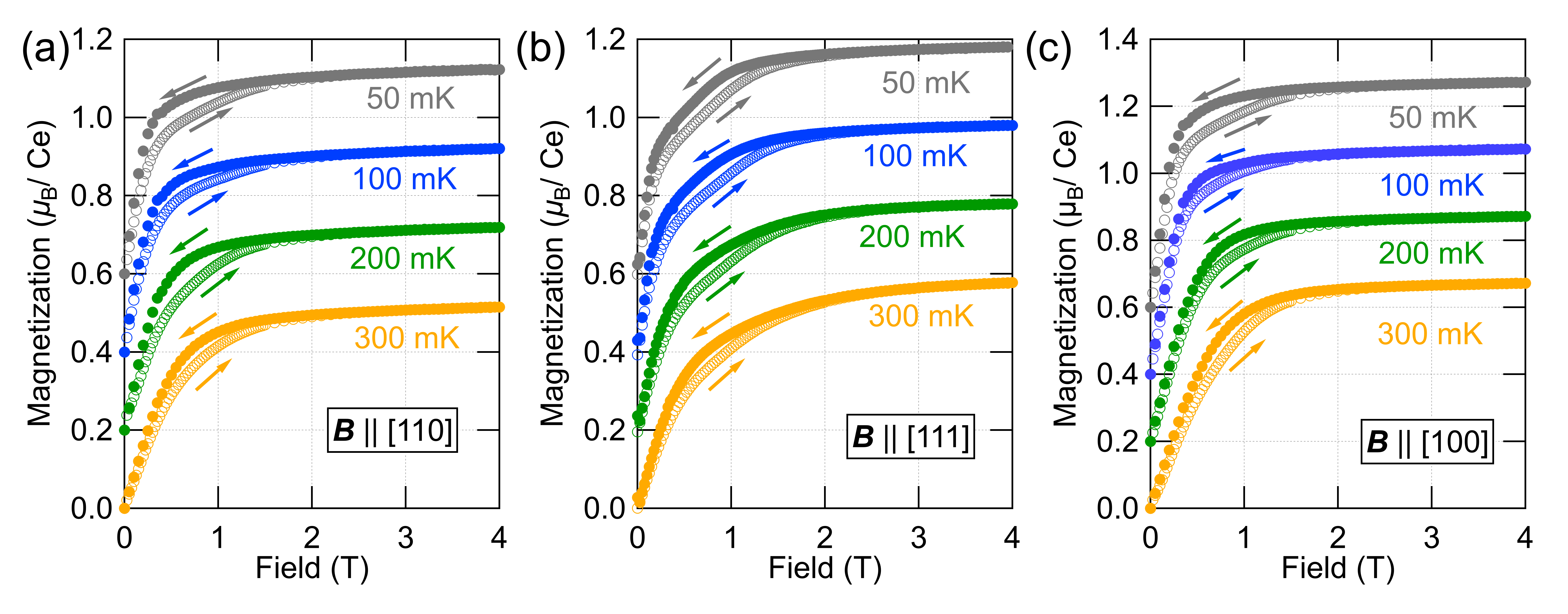}
    \hfill
    \caption{\label{fig:MH_below300mK}
    The field dependence of the magnetization under $\mathbf{B} \parallel[110]$ (a), [111] (b), and [100] (c).
The data obtained in the field-increasing and field-decreasing processes are shown by open and filled circles, respectively.
The data are vertically shifted for clarity.
    }
\end{figure}

\begin{figure}[bht!]
    \centering
    \includegraphics[width=0.4\columnwidth]{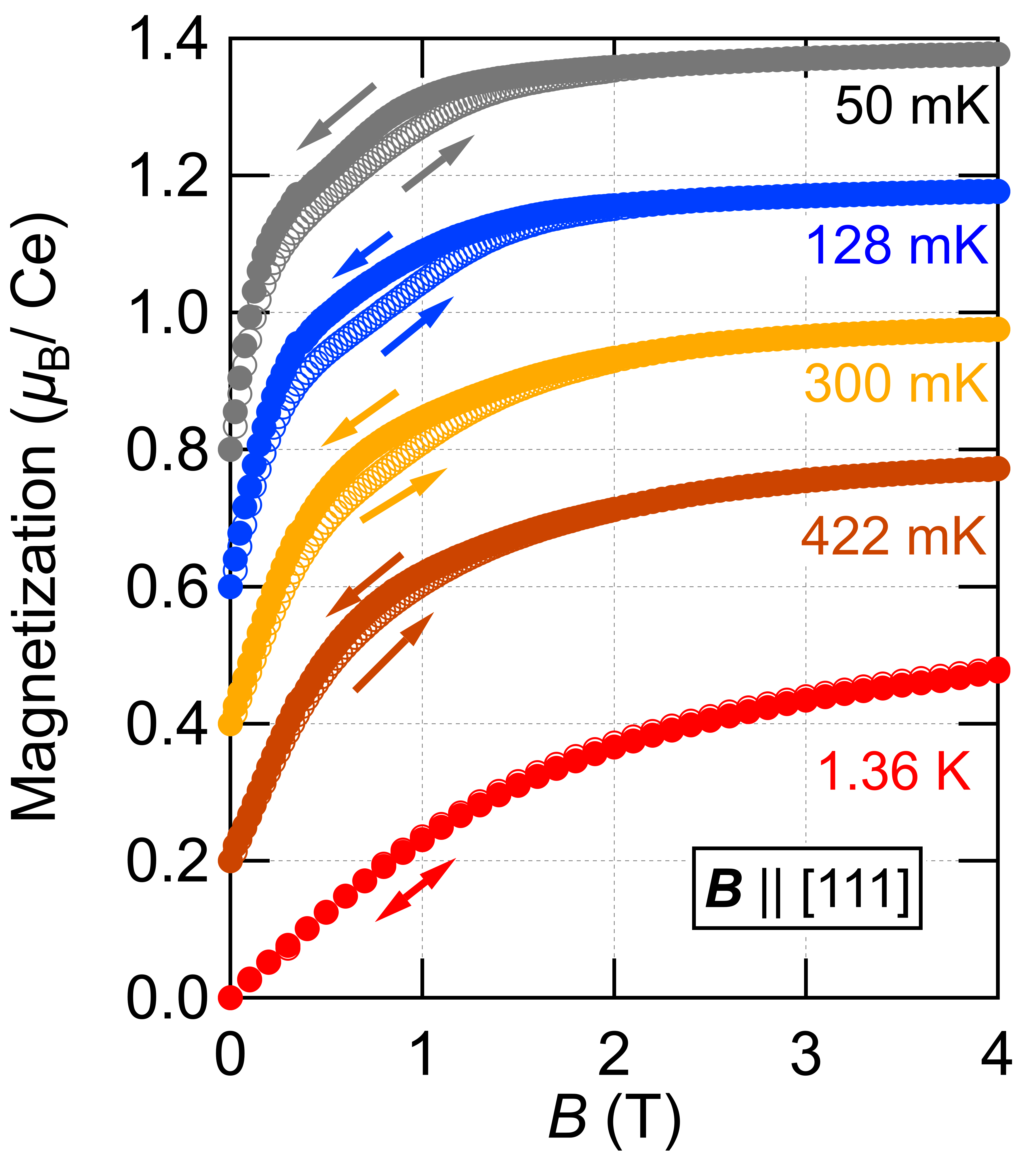}
    \caption{\label{fig:MH_50mK_1K}
    The field dependence of the magnetization under $\mathbf{B} \parallel[111]$ measured in the second run.
The data obtained in the field-increasing and field-decreasing processes are shown by open and filled circles, respectively.
The data are vertically shifted for clarity.
    }
\end{figure}

The magnetic field dependence of magnetization at lower temperatures below 300~mK is shown in Fig.~\ref{fig:MH_below300mK}.
As shown in Fig.~\ref{fig:MH_below300mK}, $M(B)$ under all three magnetic field directions starts to show a magnetic
hysteresis developing at lower temperatures, below around 2~T between the data obtained in the field-increasing and
field-decreasing processes.
In addition, two rapid slope-change features start to emerge under $\mathbf{B} \parallel [111]$
below 200~mK, near 0.35~T and 1.2~T [Fig.~\ref{fig:MH_below300mK}(b)].
The hysteresis and slope-change features are no longer visible above 300~mK, as shown in Fig.~\ref{fig:MH_50mK_1K}.

\newpage

\section{S3: Effect of relaxation time on the magnetization measurements}

\begin{figure}[htb!]
    \centering
    \includegraphics[width=0.8\columnwidth]{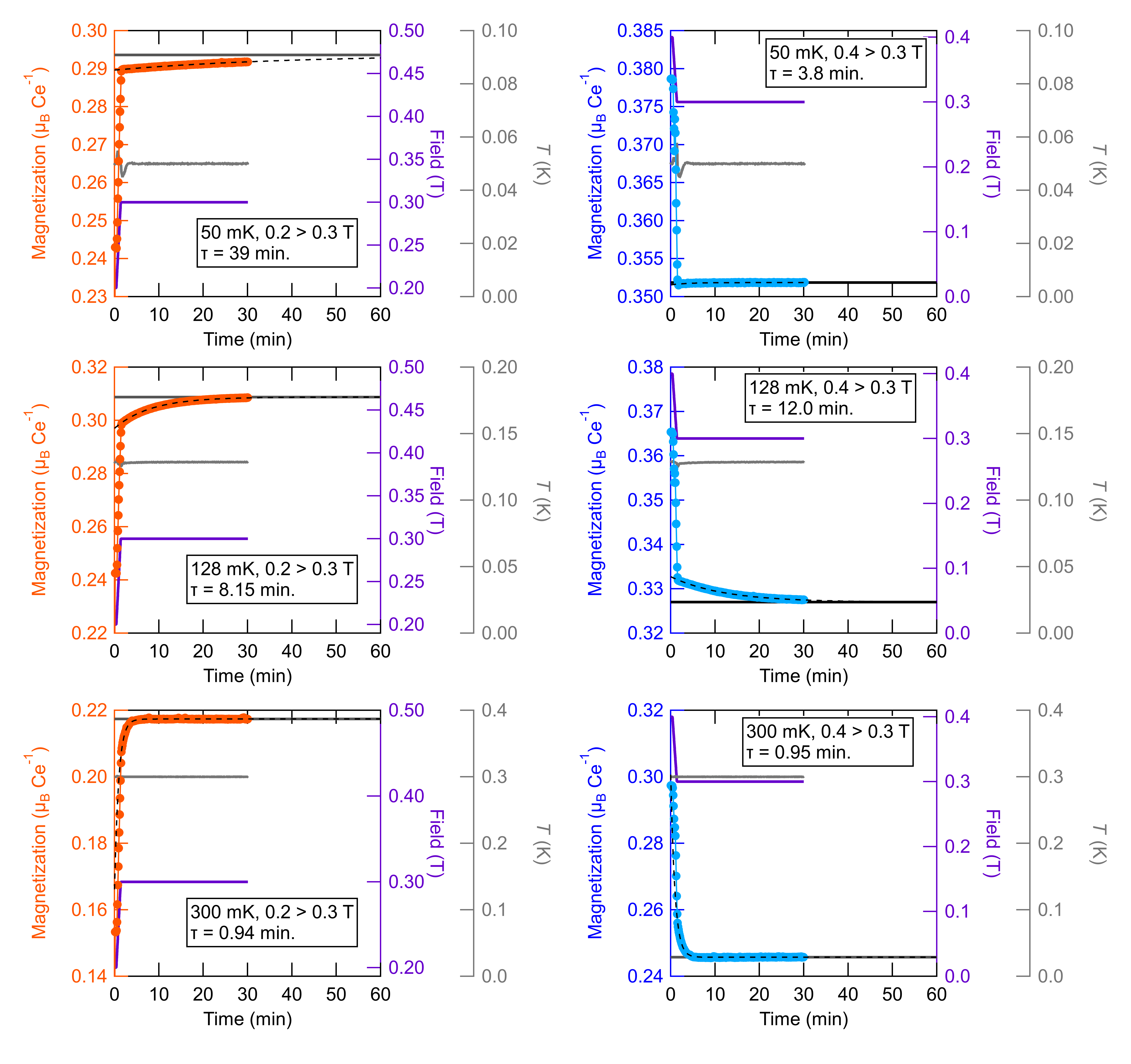}	  \hfill
    \caption{\label{fig:rlx}
    Time evolutions of the magnetization (orange (cyan) in the field-increasing (field-decreasing) measurement, left axis),
    the magnetic field applied parallel to [111] (purple, right axis), and the temperature (grey, second right axis) at
    50~mK, 128~mK, and 300~mK.
The data show the readings during the 30 minutes following the field change to 0.3~T.
    The black dashed and solid lines show an exponential fit to the magnetization data and the saturation value determined
    by the fit, respectively.
The time constant ($\tau$) determined by the fit is indicated in each panel.
    }
\end{figure}

To determine the magnetization value in the magnetization measurements, we usually wait until the rate of change of
magnetization falls below a certain threshold (typically less than 10 minutes), after the magnetic field and temperature have
stabilized.
To check whether the magnetization data reached equilibrium, we performed additional magnetization measurements
for $\mathbf{B}\parallel [111]$ by waiting 30 minutes for each measurement to compare the relaxation of the magnetization at
different temperatures and in the field-increasing and field-decreasing processes (Fig.~\ref{fig:rlx}).
Figure~\ref{fig:rlx} shows a typical time dependence of the magnetization at 0.3~T recorded in these additional measurements.
To determine the time constant $\tau$ and saturation value $M_\textrm{sat}$, we fit the time evolution to
\[
M(t)=M_\textrm{sat}+A\exp[-(t-t_0)/\tau],
\]
as shown by the dashed lines in Fig.~\ref{fig:rlx}.
We find that the magnetization reaches $M_\textrm{sat}$ (the solid lines in Fig.~\ref{fig:rlx}) within the measurement time, except for the slight deviation in the field-increasing process at 50~mK.

\begin{figure}[bht!]
    \centering
    \includegraphics[width=0.4\columnwidth]{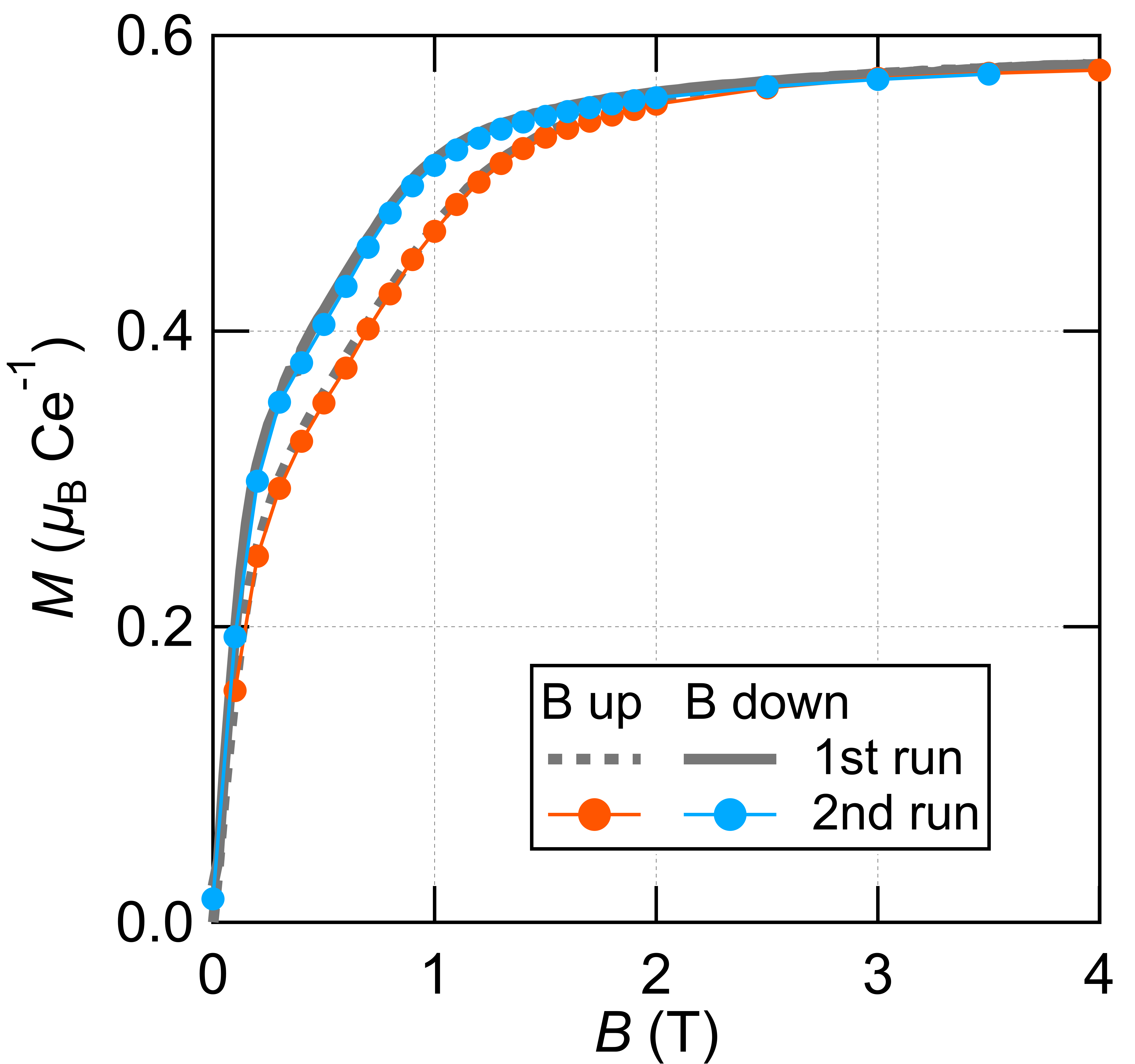}
    \caption{\label{fig:MH_50mK_rlx}
    Comparison of the field dependence of the magnetization at 50~mK under $\mathbf{B} \parallel [111]$.
The grey lines show the data obtained using the standard measurement procedure (1st run).
The circles show the saturation value determined by the exponential fitting shown in Fig.~\ref{fig:rlx} (2nd run).}
\end{figure}

Figure~\ref{fig:MH_50mK_rlx} compares the data obtained using the standard measurement procedure (the grey lines, the same data shown in the main text) and the saturation value determined by an exponential fitting of the data (circles).
As shown in Fig.~\ref{fig:MH_50mK_rlx}, the difference between the data obtained using the standard measurement procedure and the saturation value is negligible, showing that the slope-change features are not artifacts of the measurement procedure.

The long relaxation time observed in the field-increasing process at 50~mK might be relevant to the magnetic hysteresis between the field-increasing and field-decreasing measurements (Fig.~\ref{fig:MH_below300mK}).
A similar magnetic hysteresis is also observed in CSI compounds\,\cite{Ref26_Sakakibara2003LiquidGasTransition,Ref28_Snyder2004SpinFreezing,Ref29_Matsuhira2011SpinDynamics,Ref30_Giblin2011MonopoleCurrents}, which is understood in terms of the massive degeneracy of the 2-in-2-out ice states.
Since an increase in the magnetization from the zero-field degenerate state requires coherent precessions of correlated spins
under the ice rule, the initial magnetization is inhibited more strongly at lower temperatures because of fewer creations and
annihilations of magnetic monopoles\,\cite{Ref29_Matsuhira2011SpinDynamics,Ref30_Giblin2011MonopoleCurrents,Ref31_Pomaranski2013PaulingEntropy,Ref32_Paulsen2014FarFromEquilibrium,Ref33_Kassner2015SupercooledSpinLiquid},
which is also the case of Ce$_2$Hf$_2$O$_7$ due to the local Ising anisotropy.
Therefore, the magnetization in the field-increasing process may get stuck before reaching the final state, resulting in a smaller magnetization than that obtained in the field-decreasing process.

By contrast, during demagnetization the spins evolve toward a more degenerate state, making it easier to reach the equilibrium value.
We thus focus on the magnetization data obtained in the field-decreasing process in the main text.

We note that the relaxation time becomes much longer in $\mathbf{B} \parallel [100]$ than that in $\mathbf{B} \parallel [111]$.
This anisotropic relaxation time does not affect our discussion in the main text, because the difference between the data
obtained using the standard measurement procedure and the saturation value determined by fits for $\mathbf{B} \parallel [100]$
is also negligible (smaller than the symbol size of Fig.~\ref{fig:1}(b) in the main text).

\section{S4: Reproducibility of the magnetization data}

\begin{figure}[bht!]
    \centering
    \includegraphics[width=0.8\columnwidth]{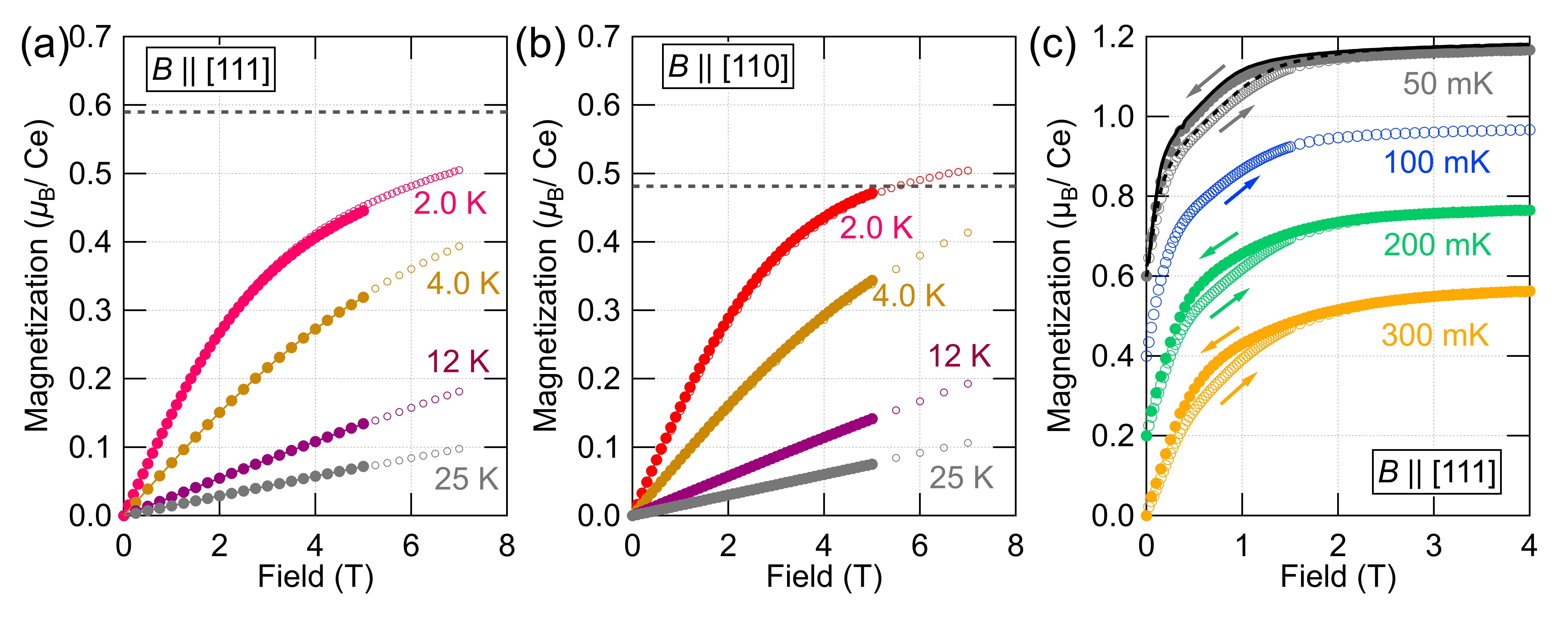}
    \caption{\label{fig:sampleB}
    (a,b) Comparison of the magnetic field dependence of the magnetization of sample B (filled symbols) and that of sample A
    (open symbols) measured under $\mathbf{B} \parallel [111]$ (a) and $\mathbf{B} \parallel [110]$ (b).
    (c) Comparison of the magnetic field dependence of the magnetization of sample B below 300~mK under
    $\mathbf{B} \parallel [111]$.
The data are vertically shifted for clarity.
The open and filled symbols show the data
    obtained in the field-increasing and field-decreasing processes, respectively.
The sample-A data at 50~mK obtained in the
    field-increasing and field-decreasing processes are shown for comparison as the dashed and solid lines, respectively.
    The demagnetization data of sample B at 100~mK are unavailable because of experimental issues.
    }
\end{figure}

The reproducibility of the magnetization data was confirmed by measurements on sample B.
As shown in Fig.~\ref{fig:sampleB}, the field dependence of the magnetization measured in sample B is virtually the same as that of sample A.

\newpage

\section{S5: Additional magnetostriction data}
\begin{figure}[hbt!]
    \centering
    \includegraphics[width=0.8\columnwidth]{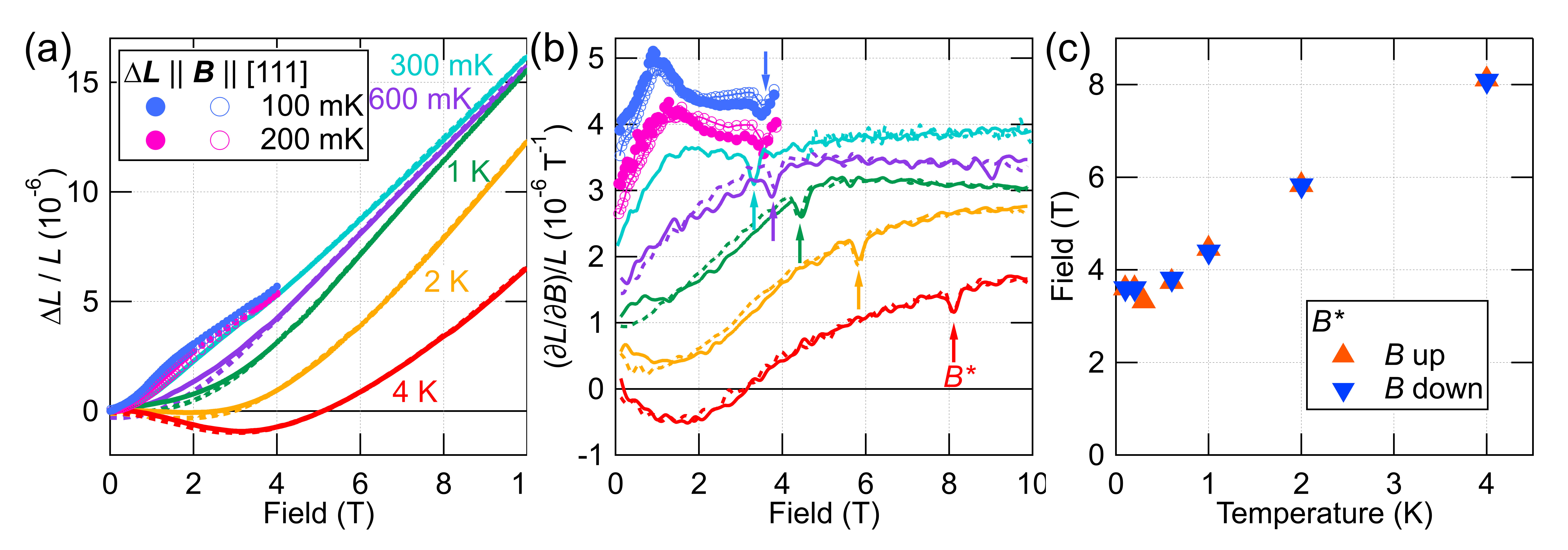}
    \caption{\label{fig:lambda}
    (a) The magnetic field dependence of the magnetostriction measured along [111] under $\mathbf{B} \parallel [111]$.
    The data above 300~mK were taken by continuously sweeping the magnetic field, whereas the data at 100~mK and 200~mK were
    taken at fixed magnetic fields.
The data obtained in the field-increasing (field-decreasing) process are shown by filled
    (open) circles for the step measurements and by solid (dashed) lines for the continuous measurements.
    (b) The field dependence of the normalized field derivative of the magnetostriction data shown using the same colors as in
    (a).
The arrows indicate the position of the dip ($B^*$).
The data are vertically shifted for clarity.
    (c) The field dependence of $B^*$ observed in the field-increasing (up triangles) and field-decreasing (down triangles)
    processes.
    }
\end{figure}

Figure~\ref{fig:lambda} (a) shows the field dependence of the magnetostriction measured along the [111] axis under $\mathbf{B} \parallel [111]$ at various temperatures.
As shown in Fig.~\ref{fig:lambda}(a), the magnetostriction at 4~K decreases with increasing field up to around 3~T, then increases above 3~T.
At higher fields, the magnetostriction becomes positive above approximately 5~T.
This positive magnetostriction becomes dominant at lower temperatures ($T\le1$~K).
Magnetic hysteresis is also observed to develop with decreasing temperature.
Below around 300~mK, a convex field dependence of the magnetostriction appears at low fields.
A linear field dependence is observed above the saturation field of the magnetization (above 4~T below 300~mK), which is due
to the local magnetic anisotropy of this compound, as observed in other spin ice
materials\,\cite{Doerr_magnetostriction_2018,stoter_PRB_2020,tang_PRB_2024}.
The linear dependence at high fields is also expected in the theoretical analysis of the magnetostriction\,\cite{Ref7_Patri2020Magnetostriction}.
We therefore subtract this linear component in Fig.~\ref{fig:1}(d) in the main text, to extract the component in the magnetostriction that corresponds to the field dependence of the magnetization.

Figure~\ref{fig:lambda} (b) shows the normalized field derivative of the magnetostriction data.
As shown in Fig.~\ref{fig:lambda}(b), a sharp dip is observed in its field derivative at around 3.8~T below 300~mK, which is
shifted to the higher fields at higher temperatures.
The field dependence of this high-field anomaly at different temperatures
is summarized in Fig.~\ref{fig:lambda}(c).
Since the magnetization is already saturated in this field range below 300~mK,
this high-field anomaly is unlikely to originate from the low-field KSL transitions discussed in the main text and may instead
reflect a lattice or high-field magnetoelastic effect.

\section{S6: Classical Monte Carlo and exact diagonalization calculations}

We calculate the field dependence of the magnetization per Ce ion,
\[
M = \frac{\mu_B g_z}{N}\sum_i
(\hat{\mathbf z}_i\cdot\hat{\mathbf B})\langle S_i^z\rangle,
\]
using the symmetry-allowed nearest-neighbor Hamiltonian for dipole-octupole pyrochlores~\cite{Ref5_Huang2014DipolarOctupolar}, given by

\begin{align}
    \mathcal{H}=\sum_{\langle ij\rangle}
    \left\lbrack J_{x}S_{i}^{x}S_{j}^{x} + J_{y}S_{i}^{y}S_{j}^{y} + J_{z}S_{i}^{z}S_{j}^{z} + J_{xz}\left( S_{i}^{x}S_{j}^{z} + S_{i}^{z}S_{j}^{x} \right) \right\rbrack - \mu_B g_{z}\sum_{i}{\left( \mathbf{B} \cdot {\widehat{z}}_{i} \right)S_{i}^{z}}.
\end{align}

This Hamiltonian can be transformed to the XYZ Hamiltonian by
eliminating the \(J_{xz}\) term by a global pseudospin rotation~\cite{Ref5_Huang2014DipolarOctupolar} as

\begin{align}
\mathcal H
=&\sum_{\langle ij\rangle}
\left[
J_{\widetilde{x}}S_i^{\widetilde{x}}S_j^{\widetilde{x}}
+J_{\widetilde{y}}S_i^{\widetilde{y}}S_j^{\widetilde{y}}
+J_{\widetilde{z}}S_i^{\widetilde{z}}S_j^{\widetilde{z}}
\right] \nonumber\\
&-\mu_B g_z\sum_i B_i^z
\left(
S_i^{\widetilde{z}}\cos\theta
+S_i^{\widetilde{x}}\sin\theta
\right),
\end{align}

where \(B_i^{z}=\mathbf B\cdot \hat z_i\), and
\begin{gather}
\theta = \frac{1}{2}\tan^{-1}\left( \frac{2J_{xz}}{J_{x} - J_{z}} \right),\\
J_{\widetilde{x}} = \frac{1}{2}\left( J_{x} + J_{z} + \sqrt{4J_{xz}^{2} + \left( J_{x} - J_{z} \right)^{2}} \right),\\
J_{\widetilde{y}} = J_{y},\\
J_{\widetilde{z}} = \frac{1}{2}\left( J_{x} + J_{z} - \sqrt{4J_{xz}^{2} + \left( J_{x} - J_{z} \right)^{2}} \right),
\end{gather}
or equivalently,
\begin{gather}
    J_{x} = J_{\widetilde{x}}\cos^{2}\theta + J_{\widetilde{z}}\sin^{2}\theta,\\
    J_{z} = J_{\widetilde{x}}\sin^{2}\theta + J_{\widetilde{z}}\cos^{2}\theta,\\
    J_{xz} = \frac{1}{2}\sin(2\theta)\left(J_{\widetilde{x}}-J_{\widetilde{z}}\right).
\end{gather}
We use the $g$ factor \(g_{z} = 2.36\) determined by the saturated
moment \(M_{\parallel} = 1.18\mu_{B}\).

As discussed in the main text, the numerical linked cluster calculations of the heat capacity have been performed to fit the heat capacity data obtained in a high-quality single crystal of Ce$_2$Hf$_2$O$_7$~\cite{Ref16_Smith2025TwoPeakHeatCapacity}.
This fit, performed in the parameter space spanned by $J_\pm / J_a = -(J_b + J_c)/(4 J_a)$ and
$J_{\pm\pm}/J_a = (J_b - J_c)/(4 J_a)$, finds minimum points of the goodness-of-fit parameter
$\langle \delta^2 / \epsilon^2 \rangle_C$ at the white circle in the QSI region and the grey circle in the ordered region
(Fig.~\ref{fig:CMC_fit}(a) in the main text).
We then select 30 points in each of the QSI and ordered regions from the best-fit point while maintaining a constant distance between them.
The parameter sets for the points in the QSI and ordered regions are listed in Table~\ref{tab:parameter_QSI} and Table~\ref{tab:parameter_ORD}, respectively.

\begin{table}[htbp]
\centering
\caption{30 Parameter points shown in Fig.~\ref{fig:CMC_fit}(a) selected in the QSI region from the smallest goodness-of-fit parameter $\langle \delta^2 / \epsilon^2 \rangle_C$ estimated in Ref.\;\cite{Ref16_Smith2025TwoPeakHeatCapacity}.}
\label{tab:parameter_QSI}
\begin{ruledtabular}
\begin{tabular}{c c c c c c c}
Point No. & $J_{\pm}/J_a$ & $J_{\pm\pm}/J_a$ & $\langle \delta^2 / \epsilon^2 \rangle_C$ & $J_a$ & $J_b$ & $J_c$ \\
1  & -0.1275 & 0.0825 & 1226 & 0.0496 & 0.0208 & 0.0045 \\
2  & -0.1350 & 0.0900 & 1255 & 0.0489 & 0.0220 & 0.0044 \\
3  & -0.1275 & 0.0675 & 1268 & 0.0497 & 0.0194 & 0.0060 \\
4  & -0.1200 & 0.0750 & 1297 & 0.0503 & 0.0196 & 0.0045 \\
5  & -0.1350 & 0.0750 & 1307 & 0.0489 & 0.0205 & 0.0059 \\
6  & -0.1200 & 0.0600 & 1327 & 0.0503 & 0.0181 & 0.0060 \\
7  & -0.1275 & 0.0975 & 1352 & 0.0495 & 0.0223 & 0.0030 \\
8  & -0.1425 & 0.0975 & 1366 & 0.0481 & 0.0231 & 0.0043 \\
9  & -0.1350 & 0.1050 & 1385 & 0.0488 & 0.0234 & 0.0029 \\
10 & -0.1275 & 0.0525 & 1405 & 0.0497 & 0.0179 & 0.0075 \\
11 & -0.1200 & 0.0900 & 1420 & 0.0503 & 0.0211 & 0.0030 \\
12 & -0.1425 & 0.0825 & 1425 & 0.0482 & 0.0217 & 0.0058 \\
13 & -0.1200 & 0.0450 & 1437 & 0.0503 & 0.0166 & 0.0075 \\
14 & -0.1350 & 0.0600 & 1469 & 0.0490 & 0.0191 & 0.0074 \\
15 & -0.1125 & 0.0675 & 1483 & 0.0511 & 0.0184 & 0.0046 \\
16 & -0.1125 & 0.0525 & 1500 & 0.0512 & 0.0169 & 0.0061 \\
17 & -0.1425 & 0.1125 & 1502 & 0.0480 & 0.0245 & 0.0029 \\
18 & -0.1500 & 0.1050 & 1538 & 0.0474 & 0.0242 & 0.0043 \\
19 & -0.1200 & 0.0300 & 1565 & 0.0506 & 0.0152 & 0.0091 \\
20 & -0.1500 & 0.0900 & 1569 & 0.0475 & 0.0228 & 0.0057 \\
21 & -0.1275 & 0.0375 & 1573 & 0.0498 & 0.0164 & 0.0090 \\
22 & -0.1125 & 0.0375 & 1583 & 0.0512 & 0.0154 & 0.0077 \\
23 & -0.1125 & 0.0825 & 1601 & 0.0510 & 0.0199 & 0.0031 \\
24 & -0.1425 & 0.0675 & 1607 & 0.0482 & 0.0203 & 0.0072 \\
25 & -0.1200 & 0.0150 & 1663 & 0.0506 & 0.0137 & 0.0106 \\
26 & -0.1125 & 0.0225 & 1673 & 0.0513 & 0.0138 & 0.0092 \\
27 & -0.1350 & 0.0450 & 1674 & 0.0491 & 0.0177 & 0.0088 \\
28 & -0.1575 & 0.1125 & 1675 & 0.0467 & 0.0252 & 0.0042 \\
29 & -0.1500 & 0.1200 & 1683 & 0.0473 & 0.0255 & 0.0028 \\
30 & -0.1575 & 0.0975 & 1698 & 0.0468 & 0.0239 & 0.0056 \\
\end{tabular}
\end{ruledtabular}
\end{table}

\begin{table}[htbp]
\centering
\caption{30 Parameter points shown in Fig.~\ref{fig:CMC_fit}(a) selected in the ordered region from the smallest goodness-of-fit parameter $\langle \delta^2 / \epsilon^2 \rangle_C$ estimated in Ref.\;\cite{Ref16_Smith2025TwoPeakHeatCapacity}.}
\label{tab:parameter_ORD}
\begin{ruledtabular}
\begin{tabular}{c c c c c c c}
Point No. & $J_{\pm}/J_a$ & $J_{\pm\pm}/J_a$ & $\langle \delta^2 / \epsilon^2 \rangle_C$ & $J_a$ & $J_b$ & $J_c$ \\
1 & 0.0525 & 0.1275 & 1787 & 0.0512 & 0.0077 & -0.0184 \\
2 & 0.0600 & 0.1050 & 1790 & 0.0520 & 0.0047 & -0.0171 \\
3 & 0.0525 & 0.1125 & 1960 & 0.0522 & 0.0063 & -0.0172 \\
4 & 0.0450 & 0.1350 & 2053 & 0.0514 & 0.0092 & -0.0185 \\
5 & 0.0600 & 0.1200 & 2066 & 0.0510 & 0.0061 & -0.0184 \\
6 & 0.0675 & 0.0975 & 2125 & 0.0516 & 0.0031 & -0.0170 \\
7 & 0.0600 & 0.0900 & 2278 & 0.0528 & 0.0032 & -0.0159 \\
8 & 0.0675 & 0.0825 & 2422 & 0.0524 & 0.0016 & -0.0157 \\
9 & 0.0525 & 0.0975 & 2511 & 0.0532 & 0.0048 & -0.0159 \\
10 & 0.0450 & 0.1500 & 2545 & 0.0503 & 0.0106 & -0.0196 \\
11 & 0.0450 & 0.1200 & 2605 & 0.0525 & 0.0079 & -0.0173 \\
12 & 0.0525 & 0.1425 & 2665 & 0.0501 & 0.0090 & -0.0196 \\
13 & 0.0675 & 0.1125 & 2831 & 0.0506 & 0.0046 & -0.0182 \\
14 & 0.0600 & 0.0750 & 2831 & 0.0536 & 0.0016 & -0.0145 \\
15 & 0.0675 & 0.0675 & 2888 & 0.0531 & 0.0000 & -0.0143 \\
16 & 0.0375 & 0.1575 & 2902 & 0.0503 & 0.0121 & -0.0196 \\
17 & 0.0375 & 0.1425 & 2922 & 0.0515 & 0.0108 & -0.0185 \\
18 & 0.0750 & 0.0750 & 2954 & 0.0520 & 0.0000 & -0.0156 \\
19 & 0.0750 & 0.0900 & 2958 & 0.0512 & 0.0015 & -0.0169 \\
20 & 0.0525 & 0.0825 & 3110 & 0.0540 & 0.0032 & -0.0146 \\
21 & 0.0450 & 0.1050 & 3111 & 0.0534 & 0.0064 & -0.0160 \\
22 & 0.0450 & 0.1650 & 3186 & 0.0491 & 0.0118 & -0.0206 \\
23 & 0.0375 & 0.1725 & 3222 & 0.0492 & 0.0133 & -0.0207 \\
24 & 0.0600 & 0.1350 & 3232 & 0.0498 & 0.0075 & -0.0194 \\
25 & 0.0750 & 0.0600 & 3296 & 0.0526 & -0.0016 & -0.0142 \\
26 & 0.0375 & 0.1875 & 3370 & 0.0479 & 0.0144 & -0.0216 \\
27 & 0.0675 & 0.0525 & 3431 & 0.0537 & -0.0016 & -0.0129 \\
28 & 0.0300 & 0.1950 & 3488 & 0.0480 & 0.0158 & -0.0216 \\
29 & 0.0600 & 0.0600 & 3494 & 0.0542 & 0.0000 & -0.0130 \\
30 & 0.0300 & 0.2100 & 3556 & 0.0466 & 0.0168 & -0.0223 \\
\end{tabular}
\end{ruledtabular}
\end{table}

In these parameter sets, $\{J_a, J_b, J_c\}$ is a permutation of $\{J_{\tilde{x}}, J_{\tilde{y}}, J_{\tilde{z}}\}$.
Note that shifting \(\theta\) to \(\theta + \pi/2\) swaps
\(J_{\tilde{x}}\) and \(J_{\tilde{z}}\).
Therefore, there are
three non-equivalent permutations as
$\{J_{\tilde{x}}, J_{\tilde{y}}, J_{\tilde{z}}\} = \{J_a, J_b, J_c\}$, $\{J_a, J_c, J_b\}$, and $\{J_b, J_a, J_c\}$.

In order to find the spin interaction parameters that reproduce our magnetization data, we performed large-scale classical Monte Carlo (CMC) simulations based on the
heat-bath algorithm for a system with the linear system size \(L = 12\)
(6912 spins in total) under periodic boundary conditions.
For each of the 60 parameter points and each of the three non-equivalent permutations, we calculated the field dependence of the magnetization by CMC while scanning \(0\le \theta \le \pi\).
We then estimated the chi-square value ($\chi^2$) between  the magnetization curve obtained by the CMC simulation and the experimental data obtained in the field-decreasing process under $\mathbf{B}\parallel [100]$ and [111].
The CMC calculations performed at 25~mK were used for this estimation because they showed
good convergence.
The data measured under \(\mathbf{B} \parallel [110]\) were
excluded because this field direction is highly sensitive to misalignment~\cite{Ref24_Bhardwaj2025ThermodynamicsCe2Hf2O7}.

\begin{figure}[th!]
    \centering
    \includegraphics[width=\textwidth]{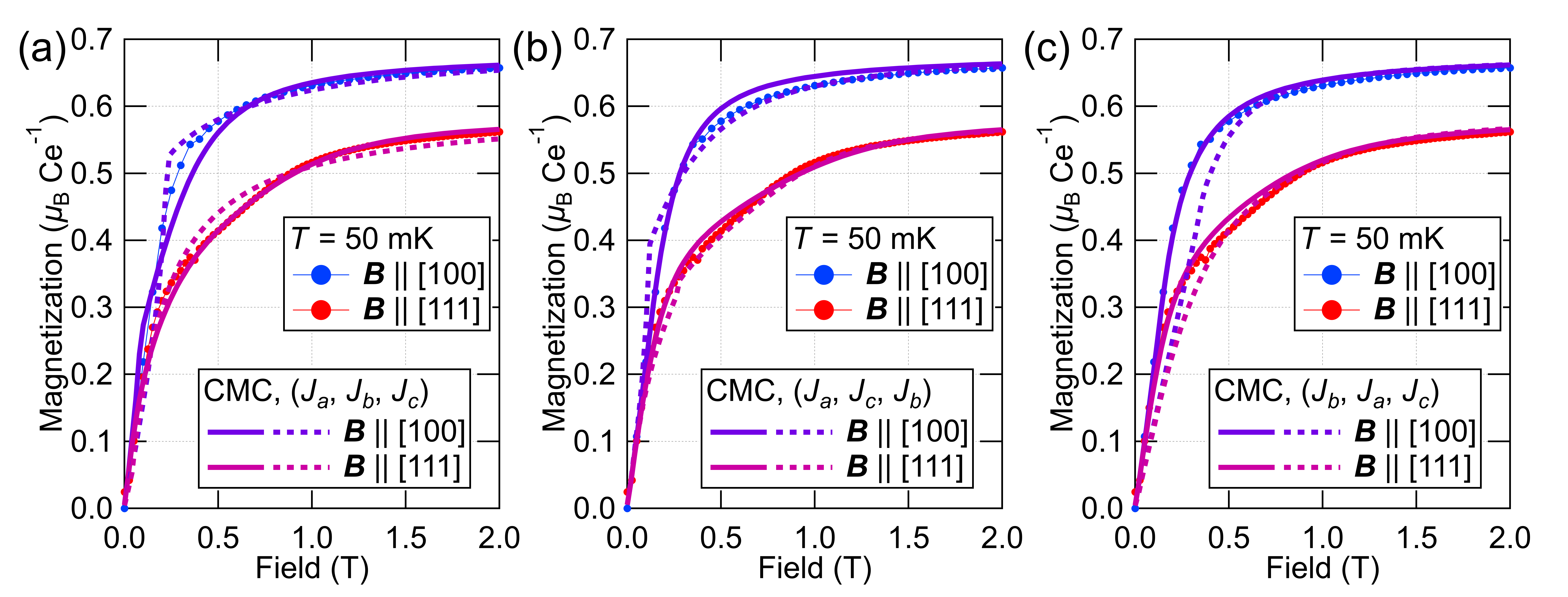}
    \caption{\label{fig:extended_sim_MC} Comparison of the magnetization curves under $\mathbf{B}\parallel [100]$ and [111]
obtained by the measurements (circles) and the classical Monte Carlo simulations calculated from the best parameter set
listed in Table~\ref{tab:fit_results} for the permutation of $\{J_a, J_b, J_c\}$ (a), $\{J_a, J_c, J_b\}$ (b), and
$\{J_b, J_a, J_c\}$ (c).
The simulation data obtained in the QSI (ordered) region are shown by the solid (dashed) line.}
\end{figure}

The distribution of $\chi^2$ is summarized in Fig.~\ref{fig:CMC_fit}(b).
As shown in Fig.~\ref{fig:CMC_fit}(b), the parameter sets in the QSI region (the top panels in
Fig.~\ref{fig:CMC_fit}(b)) result in smaller $\chi^2$ values than those in the ordered region (the bottom panels) for all the
permutations, which is consistent with the smaller goodness-of-fit parameter $\langle \delta^2 / \epsilon^2 \rangle_C$ in
the QSI region~\cite{Ref16_Smith2025TwoPeakHeatCapacity}.
The best combination of the parameter point and $\theta$ for each permutation is marked by the white circles and is listed in Table~\ref{tab:fit_results}.
The magnetization curves obtained by these best parameters are shown with the experimental data in Fig.~\ref{fig:extended_sim_MC}.
As shown in Table~\ref{tab:fit_results},
the dominant interaction lies either in the $J_x$ or $J_y$ channel, with the best overall fit dominated by the $J_y$ channel associated with the octupolar component.
This extended CMC analysis supports the representative Hamiltonian used in the main text, which features strong transverse, octupolar exchange and lies in the QSI-compatible regime.

\begin{table}[ht]
\centering
\caption{The best-fit parameters obtained from the CMC simulations for the parameter points shown in Fig.~\ref{fig:CMC_fit}(b) in the main text.
The dominant interaction for each permutation is highlighted in bold.}
\label{tab:fit_results}
\begin{ruledtabular}
\begin{tabular}{l l c c c c c c c}
Phase & Permutation & Point No. & $\theta/\pi$ & $\chi^2$ & $J_x$ (meV) & $J_y$ (meV) & $J_z$ (meV) & $J_{xz}$ (meV) \\
QSI     & abc & 25 & 0.08 & $2.534 \times 10^{-4}$ & \textbf{0.0485} & 0.0140 & 0.0135 & 0.0096 \\
QSI     & acb & 27 & 0.04 & $1.008 \times 10^{-4}$ & \textbf{0.0485} & 0.0090 & 0.0185 & 0.0039 \\
QSI     & bac & 29 & 0.24 & $0.783 \times 10^{-4}$ & 0.0152 & \textbf{0.0470} & 0.0138 & -0.0115 \\
Ordered & abc & 58 & 0.28 & $2.617 \times 10^{-4}$ & 0.0064 & 0.0160 & 0.0196 & \textbf{0.0344} \\
Ordered & acb & 34 & 0.84 & $1.769 \times 10^{-4}$ & \textbf{0.0413} & -0.0180 & 0.0187 & -0.0177 \\
Ordered & bac & 41 & 0.48 & $17.13 \times 10^{-4}$ & -0.0169 & \textbf{0.0520} & 0.0079 & 0.0016 \\
\end{tabular}
\end{ruledtabular}
\end{table}

\begin{figure}[th!]
    \centering
    \includegraphics[width=\textwidth]{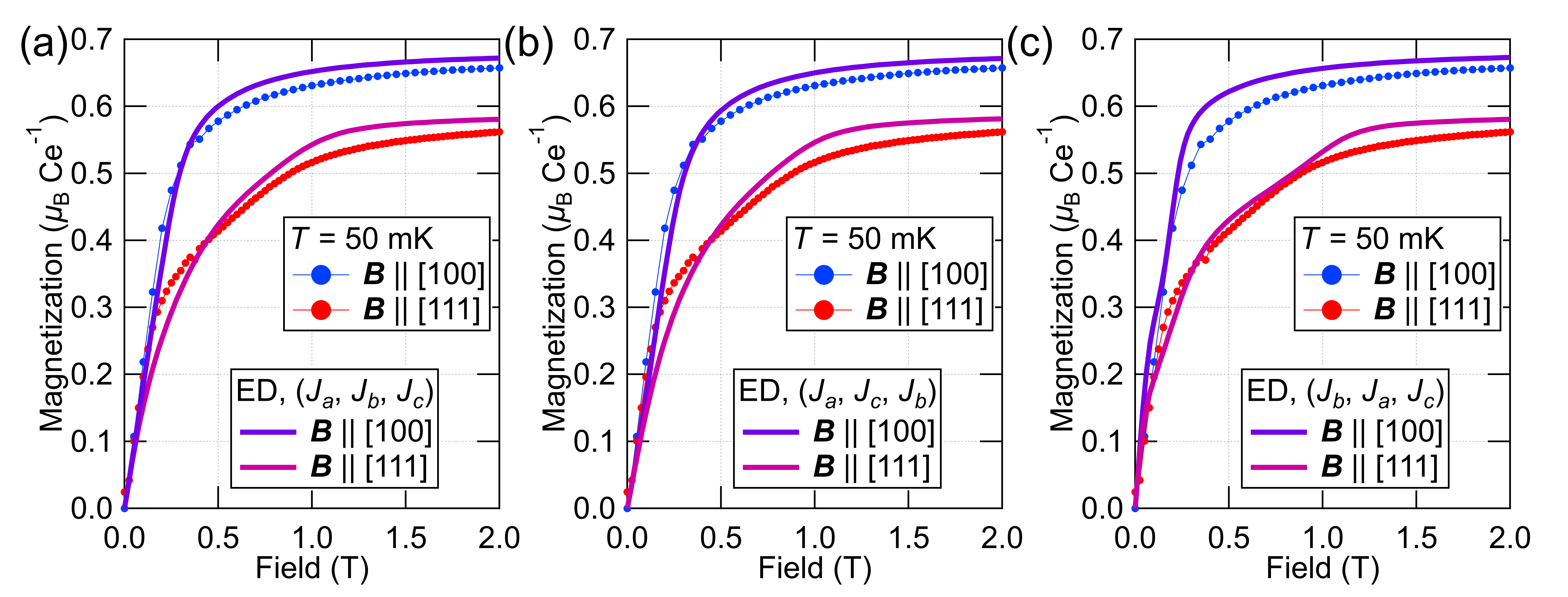}
    \caption{\label{fig:extended_sim_ED} Comparison of the magnetization curves under $\mathbf{B}\parallel [100]$ and [111]
obtained by the measurements (circles) and the exact diagonalization calculated from the best parameter set in the QSI region
listed in Table~\ref{tab:fit_results} for the permutation of $\{J_a, J_b, J_c\}$ (a), $\{J_a, J_c, J_b\}$ (b), and
$\{J_b, J_a, J_c\}$ (c).}
\end{figure}

We further carried out finite-temperature exact diagonalization (ED) to calculate the field-dependent magnetization for the QSI-region parameter sets.
The ED calculations were performed on a 16-site cluster using the Lanczos algorithm, obtaining the lowest 200 eigenvalues, which were sufficient to converge the magnetization.
All the magnetization curves obtained by the ED calculations are shown in Fig.~\ref{fig:extended_sim_ED}.

As shown in Fig.~\ref{fig:extended_sim_ED}, the ED calculations reproduce the magnetization curve well, especially the two slope-change anomalies for $\mathbf{B}\parallel [111]$ shown in Fig.~\ref{fig:1}(c) in the main text.
By contrast, the deviation from the experimental data is larger in ED than in CMC.
Comparing the CMC and quantum ED calculations, CMC accesses much larger clusters and therefore better approximates the
thermodynamic limit and extended correlations, but it neglects quantum fluctuations.
ED, in contrast, includes local quantum
fluctuations exactly on the finite cluster at the experimental temperatures via Boltzmann weighting of the spectrum, but its
principal systematic error is finite-size and boundary-condition bias.
We therefore interpret the two approaches as
complementary: both place the system in the same qualitative regime of the dipole-octupole model, while their residual
differences quantify combined finite-size, quantum-classical, and model uncertainties.

\end{document}